\documentclass[twocolumn]{aastex631}

\usepackage{hyperref}
\usepackage{xcolor}

\newcommand{\mytilde}{\raise.19ex\hbox{$\scriptstyle\sim$}}
\shorttitle{Galaxy Image Restoration with Efficient Transformer}
\shortauthors{Park et al.}

\begin{document}

\title{Deeper, Sharper, Faster: Application of Efficient Transformer to  Galaxy Image Restoration}
\correspondingauthor{M. James Jee}
\email{hyosun.park@yonsei.ac.kr, mkjee@yonsei.ac.kr}

\author[0009-0007-7093-1758]{Hyosun park}
\affiliation{Department of Astronomy, Yonsei University, Seoul, Republic of Korea}

\author[0009-0005-9082-4875]{Yongsik Jo}
\affiliation{Artificial Intelligence Graduate School, UNIST, Ulsan, Republic of Korea}

\author[0000-0003-4243-8214]{Seokun Kang}
\affiliation{Artificial Intelligence Graduate School, UNIST, Ulsan, Republic of Korea}

\author[0000-0002-6571-4632]{Taehwan Kim}
\affiliation{Artificial Intelligence Graduate School, UNIST, Ulsan, Republic of Korea}

\author[0000-0002-5751-3697]{M. James Jee}
\affiliation{Department of Astronomy, Yonsei University, Seoul, Republic of Korea}
\affiliation{Department of Physics and Astronomy, University of California, Davis, CA, USA}

\begin{abstract}
The Transformer architecture has revolutionized the field of deep learning over the past several years in diverse areas, including natural language processing, code generation, image recognition, time series forecasting, etc. We propose to apply Zamir et al.'s efficient transformer to perform deconvolution and denoising to enhance astronomical images. We conducted experiments using pairs of high-quality images and their degraded versions, and our deep learning model demonstrates exceptional restoration of photometric, structural, and morphological information. When compared with the ground-truth JWST images, the enhanced versions of our HST-quality images reduce the scatter of isophotal photometry, Sersic index, and half-light radius by factors of 4.4, 3.6, and 4.7, respectively, with Pearson correlation coefficients approaching unity.
The performance is observed to degrade when input images exhibit correlated noise, point-like sources, and artifacts. We anticipate that this deep learning model will prove valuable for a number of scientific applications, including precision photometry, morphological analysis, and shear calibration.
\end{abstract}

\keywords{techniques: image processing --- galaxies: fundamental parameters --- galaxies: photometry --- galaxies: structure --- deep learning --- astronomical databases: images}

\section{Introduction} \label{sec:intro}
A deeper and sharper image of an astronomical object offers an opportunity for gaining new insights and understanding of it. One of the most notable examples is the advent of the James Webb Space Telescope (JWST), launched in 2021, which has since continuously expanded our knowledge of the universe through its discoveries, thanks to its unprecedented resolution and depth. This trend is reminiscent of a similar era three decades ago when the Hubble Space Telescope (HST) became the most powerful instrument available to us at that time.

Astronomers' ongoing efforts to enhance the depth and clarity of astronomical imaging extend beyond advancements in instrumentation to include significant developments in the algorithmic domain. Early techniques relied on straightforward Fourier deconvolution techniques ~\citep[e.g.,][]{simkin1974, joneswykes1989J}. A primary challenge with this approach is noise amplification, and linear filtering was suggested to remedy the frequency-dependent signal-to-noise ratio issue ~\citep[e.g.,][]{tikhonov1987}. However, the filtering method provided only band-limited results. With the development of computer technologies, more computationally demanding, sophisticated approaches based on the Bayesian principle with various regularization schemes were introduced ~\citep[e.g.,][]{richardson1972, lucy1974, sheppvardi1982}. Some implementations of these Bayesian approaches were shown to outperform the early Fourier deconvolution methods. Nevertheless, the regularization could not recover both compact and extended features simultaneously well. Multi-resolution or spatially adaptive approaches were suggested to overcome this limitation~\citep[e.g.,][]{wakker1998,yan2012}.

In recent years, the emergence of deep learning has significantly impacted the general field of
image restoration and enhancement. Also, a growing number of studies are reporting notable results in astronomical contexts as well ~\citep[e.g.,][]{schawinski2017, sureau2020, Lanusse2021, akhaury2022, sweere2022}.
Deep learning algorithms, particularly convolutional neural networks \citep[CNNs;][]{krizhevsky2012imagenet}, have shown promising results in automatically learning complex features from images and effectively enhancing their depth and resolution ~\citep[e.g.,][]{zhang2017, diazbaso2019, elhakiem2021, zhang2022}. By training on large datasets of both raw and enhanced images, deep learning models can learn intricate patterns and relationships within the data, allowing for more precise and tailored image enhancement.

Deep learning techniques offer advantages over traditional methods, such as Fourier deconvolution and Bayesian approaches, by providing greater flexibility and adaptability to diverse datasets. The integration of CNNs with other deep learning architectures, such as generative adversarial networks ~\citep[GANs;][]{goodfellow2014generative} and recurrent neural networks ~\citep[RNNs;][]{williams1989learning}, has further expanded the capabilities of image enhancement ~\citep[e.g.,][]{ledig2016, schawinski2017, liu2018, tripath2018, alsaiari2019, rajeev2019, wang2020, tran2020, kalele2023}. These hybrid approaches enable the generation of highly realistic and detailed images, pushing the boundaries of what is achievable with traditional methods alone.
One of the outstanding limitations of the CNN-based model is its restricted receptive field. That is, long-range pixel correlations may not effectively be captured by the model. 
Another critical drawback is its limited adaptability to input content during inference.

The Transformer architecture \citep{vaswani2017}, which has revolutionized the field of deep learning over the past several years in diverse areas including the well-known large language models, can be considered a potent alternative to overcome these limitations of the CNN-based models. However, with its original implementation structure comprised of so-called self-attention layers, it is infeasible to apply the Transformer model to large images because the computing complexity increases quadratically with the number of pixels. 

\citet{zamir2022restormer} devised an innovative scheme to substitute the original self-attention block with the multi-Dconv transposed attention (MDTA) block. The MDTA block, implementing self-attention in the feature domain rather than the pixel domain, ensures that the complexity increases only linearly with the number of pixels, making its application to large images feasible. \citet{zamir2022restormer} demonstrated that their model, named {\tt Restormer}, attained state-of-the-art results for image deraining, single-image motion deblurring, defocus deblurring, and image denoising. However, its performance has not been evaluated in the astronomical context. 

In this paper, we propose to apply Zamir et al.'s efficient transformer to perform deconvolution and denoising to enhance astronomical images. Specifically, we investigate the feasibility of enhancing HST images to achieve the quality of JWST ones in both resolution and depth. We build our model based on \cite{zamir2022restormer}'s implementation and employ the transfer learning approach, initially pre-training the model using simplified galaxy images and then finetuning it using realistic galaxy images.

Our paper is structured as follows. In \textsection\ref{sec:Method}, we describe the overall architecture of {\tt Restormer} and the implementation details employed in the current galaxy restoration model. 
\textsection\ref{sec:data} explains how we prepare training datasets. Results are presented in \textsection\ref{sec:results}. We show the results when the model is applied to real HST images in \textsection\ref{sec:realapplication} and discuss the limitations in \textsection\ref{sec:limitation} before we conclude in \textsection\ref{sec:conclusion}.

\section{Method} \label{sec:Method}

\subsection{Overview} \label{subsec:Overview}
\begin{figure*}
    \centering
    \includegraphics[width=\textwidth]{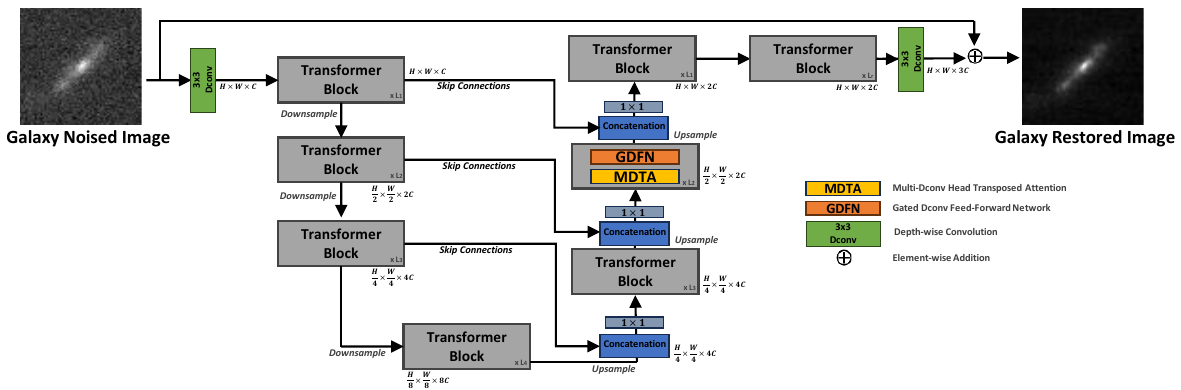}
    \caption{{\tt Restormer} architecture employed in this paper for galaxy image restoration. The architecture uses a multi-scale hierarchical design incorporating efficient Transformer blocks. The two core modules of the  Transformer block are MDTA and GDFN~\citep{zamir2022restormer}. 
    }
    \label{fig:restormer}
\end{figure*}

Throughout the paper, we use the term restoration to refer to the process that simultaneously improves resolution and reduces noise.
Our goal is to restore the HST-quality image to the JWST-quality image based on the {\tt Restormer} implementation \citep{zamir2022restormer} of the Transformer architecture \citep{vaswani2017}.
We first briefly review the encoder-decoder architecture in \textsection\ref{subsec:unet}. 
\textsection\ref{subsec:transformers} describes
the Transformer architecture including the \cite{zamir2022restormer}'s implementation. Implementation details are presented in \textsection\ref{subsec:Implementation Details}.

\subsection{Encoder-Decoder Architecture} \label{subsec:unet}
The encoder-decoder architecture allows neural networks to learn to map input data to output data in a structured and hierarchical manner. The encoder captures characterizing features from the input data and encodes them into a compressed representation, while the decoder reconstructs or generates the desired output based on this encoded representation. This architecture has been widely used in various applications, including image-to-image translation, image segmentation, language translation, and more.

U-net~\citep{ronneberger2015u} is a classic example of an encoder-decoder architecture based on CNN. It consists of a contracting path, which serves as the encoder, an expanding path, which functions as the decoder, and skip connections that link corresponding layers between the contracting and expanding paths.
In the convolutional layers of the contracting path, the dimension is reduced by increasing the number of channels to capture essential image features. The expanding path employs only low-dimensional encoded information to decrease channel numbers and increase dimensions, aiming to restore high-dimensional images. To mitigate information loss in the contracting path, the skip connection is employed to concatenate features obtained from each layer of the encoding stage with each layer of the decoding stage.

\subsection{Transformers for Image Restoration} \label{subsec:transformers}

In Transformer, the encoder consists of multiple layers of self-attention mechanisms followed by position-wise feed-forward neural networks.
``Attention" refers to a mechanism that allows models to focus on specific parts of input data while processing it. It enables the model to selectively weigh different parts of the input, giving more importance to relevant information and ignoring irrelevant or less important parts.
The key idea behind attention is to dynamically compute weights for different parts of the input data, such as words in a sentence or pixels in an image, based on their relevance to the current task.
In self-attention, each element (e.g., word or pixel) in the input sequence is compared to every other element to compute attention weights, which represent the importance of each element with respect to others. These attention weights are then used to compute a weighted sum of the input elements, resulting in an attention-based representation that highlights relevant information.

The Transformer decoder also consists of multiple layers of self-attention mechanisms, along with additional attention mechanisms over the encoder's output. The decoder predicts one element of the output sequence at a time, conditioned on the previously generated elements and the encoded representation of the input sequence.

The Transformer architecture was initially proposed and applied to the task of machine translation, which involves translating text from one language to another.
The success of the Transformer in machine translation tasks demonstrated its effectiveness in capturing long-range dependencies in sequences and handling sequential data more efficiently than traditional architectures. This breakthrough sparked widespread interest in the Transformer architecture, leading to its adoption and adaptation for various image processing tasks. 
Transformers show promising results in tasks such as image classification, object detection, semantic segmentation, and image generation, traditionally dominated by CNNs. Transformer models capture long-range pixel correlations more effectively than CNN-based models. 

However, using the Transformer model on large images becomes challenging with its original implementation, which applies self-attention layers on pixels. This is because the computational complexity escalates quadratically with the pixel count. \cite{zamir2022restormer} overcame this obstacle by substituting the original self-attention block with the MDTA block, which implements self-attention in the feature domain and makes the complexity increase only linearly with the number of pixels.
We propose to use \cite{zamir2022restormer}'s efficient Transformer {\tt Restormer} to apply deconvolution and denoising to astronomical images. We briefly describe the two core components of {\tt Restormer} in \textsection\ref{subsec:MDTA} and \textsection\ref{subsec:GDFN}. Readers are referred to \cite{zamir2022restormer} for more technical details.

\subsubsection{MDTA block} \label{subsec:MDTA}
MDTA stands as a crucial module within {\tt Restormer}. By performing self-attention in the channel dimension, MDTA calculates interactions between channels in the input feature map, creating query-key interactions. Through this process, MDTA effectively models interactions between channels in the input feature map, facilitating the learning of the global context necessary for image restoration tasks.

MDTA also employs depth-wise convolution to accentuate local context. This enables MDTA to emphasize the local context of the input image, ultimately allowing for the modeling of both global and local contexts.

\subsubsection{GDFN block} \label{subsec:GDFN}
GDFN, short for Gated-Dconv Feed-Forward Network, stands as another crucial module within {\tt Restormer}. Utilizing a gating mechanism to enhance the Feed-Forward Network, GDFN offers improved information flow, resulting in high-quality outcomes for image restoration tasks.

GDFN controls the information flow through gating layers, composed of element-wise multiplication of two linear projection layers, one of which is activated by the Gaussian Error Linear Unit (GELU) non-linearity. This allows GDFN to suppress less informative features and hierarchically transmit only valuable information. Similar to the MDTA module, GDFN employs local content mixing. Through this, GDFN emphasizes the local context of the input image, providing a more robust information flow for enhanced results in image restoration tasks.

\begin{figure*}
    \centering
    \includegraphics[width=\textwidth]{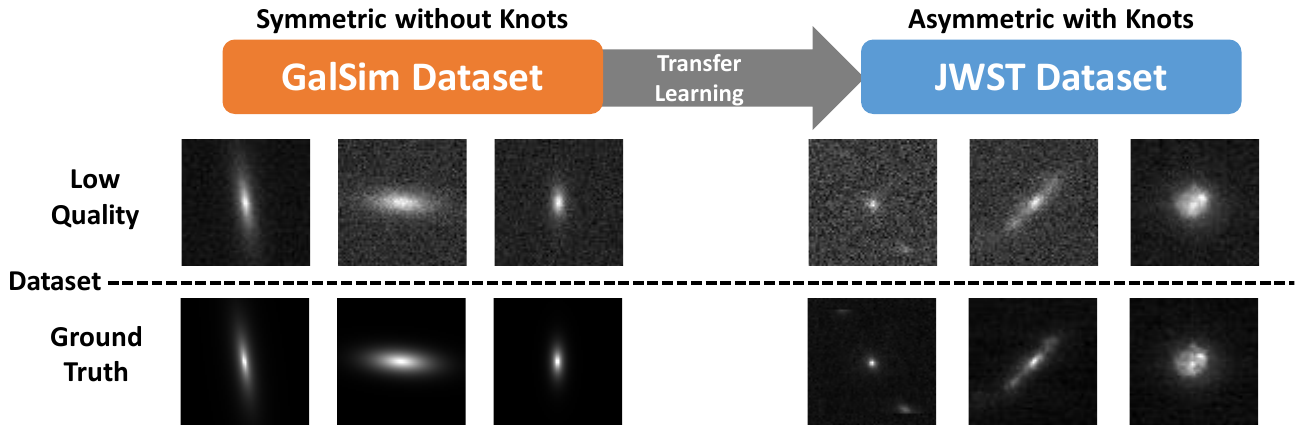}
    \caption{Galaxy image restoration through transfer learning. Initially, we pretrain our model with LQ-GT pairs of simplified galaxy images created with {\tt GalSim}.
    Then, this model is finetuned with LQ-GT pairs of
    realistic galaxy images sampled from deep JWST images. We find that this transfer learning scheme shows improved performance over the models trained with single dataset.
    }
    \label{fig:learning}
\end{figure*}

\subsection{Implementation Details} \label{subsec:Implementation Details}
In all experiments, we use the following training parameters unless mentioned otherwise. 
Our Restormer employs a 4-level encoder-decoder. From level 1 to level 4, the number of Transformer blocks is [4, 6, 6, 8], attention heads in MDTA are [1, 2, 4, 8], and the number of channels is [48, 96, 192, 384]. 
The refinement stage contains 4 blocks. The channel expansion factor in GDFN is r=2.66.
We train models with the AdamW \citep{loshchilov2018fixing} optimizer ($B1 =0.9$, $B2 =0.999$, and a weight decay of $10^{-4}$) and L1 loss for 300,000 iterations with the initial learning rate $3\times10^{-4}$ gradually reduced to $10^{-6}$ with the cosine annealing. 

Every input image is min-max normalized. 
This normalization scheme is needed because 
astronomical images span a wide dynamic range. Without the proper normalization, the training will be biased toward restoring the brightest galaxy images.
Also, we find that this standardization helps to train our model more efficiently. Finally, the min-max normalization minimizes the effects of field-dependent calibration systematics.

We use a transfer learning approach, where a model trained on one dataset is reused for another related dataset.
First, we train on the pre-training dataset (simplified galaxy images) for 150,000 iterations, followed by an additional 150,000 iterations on the finetuning dataset (realistic galaxy images). The batch size remains fixed at 64. Additionally, to compare and analyze the performance of training on individual datasets, we also conduct training separately solely based on either the pre-training or finetuning datasets.
Our inference model is publically available \footnote{\url{https://github.com/JOYONGSIK/GalaxyRestoration}} \footnote{Zenodo: \url{https://doi.org/10.5281/zenodo.11378660}} .

\section{Data \label{sec:data}}
The goal of the current study is to construct a Transformer model that enhances the quality of HST images to that of the JWST. Thus, our training data set 
consists of pairs of JWST-quality ground truth (GT) images and their degraded, low-quality (LQ) versions.
We employ a transfer-learning approach by first pre-training our model with GT-LQ pairs of simplified galaxy images and finetuning the pre-trained model with GT-LQ pairs of realistic galaxy images. 
Therefore, we prepare two sets of training datasets.

The GT image of the pre-training dataset is generated with {\tt GalSim} \citep{Rowe2015galsim} while its LQ counterpart is obtained by simply degrading the resolution of the GT image with the HST point spread function (PSF) and adding Gaussian noise. The input parameters of {\tt GalSim} should not be arbitrary but must be chosen carefully to represent the structural features of real galaxies. This is achieved by harvesting real galaxy images from the HST archive and measuring their properties.

The GT image of the finetuning dataset is obtained by collecting galaxy images from the JWST archive. Following the same procedure in the pre-training dataset generation, we create their LQ counterparts. 

We provide detailed descriptions of the HST, {\tt GalSim}, and JWST datasets in \textsection\ref{subsec:HST}, 
\textsection\ref{subsec:Galsim}, and
\textsection\ref{subsec:JWST}, respectively.
Below we briefly outline the purpose of each dataset.
\begin{itemize}
  \item{HST dataset: To define the properties of galaxies for creating synthetic images and determine the appropriate noise level for generating low-quality images. This dataset is not used in training.}
  \item{GalSim dataset: To pretrain the model on noise-free synthetic images of galaxies to learn core features.}
  \item{JWST dataset: To finetune the model using real JWST galaxy images to learn diverse forms and characteristics of galaxies not covered by the GalSim dataset.}
\end{itemize}

\subsection{HST Dataset} \label{subsec:HST}

We collect multi-band galaxy images from the following HST/ACS deep fields: the Hubble Ultra Deep Field 12 (HUDF12), HUDF-Parallel 2 (HUDFP2)\footnote{\url{https://archive.stsci.edu/prepds/hudf12/}} \citep{ellis2013, koekemoer2013}, and the Great Observatories Origins Deep Survey North and South (GOODS-N$\&$S)\footnote{\url{https://archive.stsci.edu/hlsps/hlf/}} \citep{Illingworth2016, whitaker2019} fields.
The F606W, F775W, and F814W filters are used from HUDF12. For HUDFP2, only F814W is selected. We employ F606W and F814W from GOODS-North$\&$South.

\begin{figure}[ht!]
\plotone{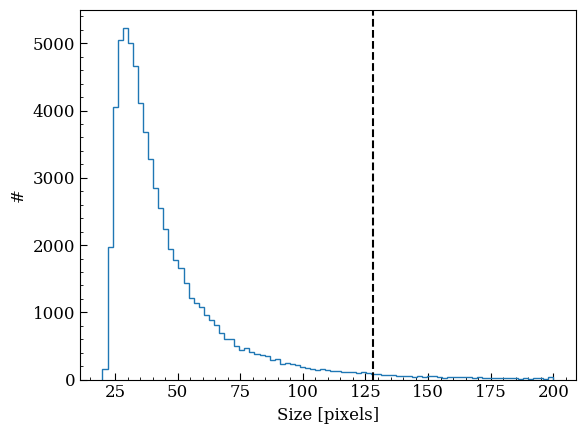}
\caption{Size distribution of the HST postage stamp images.
The postage stamp image size $w$ is determined by the following equation: $w= 2\times [4 a +5]$, where $a$ is the semi-major axis output by {\tt SExtractor} and the operation $[x]$ represents the largest integer that does not exceed $x$. We used images smaller than 128$\times$128 pixels. }
\label{fig:size}
\end{figure}

{\tt SExtractor} \citep{1996A&AS..117..393B} was used to detect sources by searching for five connected pixels
1.5 times above the rms noise level.
We cropped a $w\times w$ pixel postage-stamp image centered on each source. The image size $w$ is set to $2[4a+5]$, where $a$ is the semi-major axis output by {\tt SExtractor} and the operation $[x]$ represents the greatest integer not exceeding $x$.
We discard the source if
\begin{enumerate}
\item its image size $w$ is greater than 128 pixels,
\item either its semi-minor axis or half-light radius is smaller than the FWHM of HST/ACS PSF,
\item its S/N value is less than 20,
\item the maximum value of its neighboring sources is larger than 0.8 times its peak value, or
\item the gradient within the central $10\times10$ pixels is sharper than the HST PSF.
\end{enumerate}
The first and second criteria eliminate excessively large or small sources. The third condition removes 
excessively faint sources. The fourth condition excludes the cases where similarly bright galaxies are present within the postage stamp image. The last condition is necessary because the HST automatic cosmic ray detection/removal pipeline often fails near the bright peaks of stars or galaxies.
In addition to these criteria, we explicitly identified and removed stars using their distinct locus in the size-magnitude diagram. These conditions define our domain of interest and ensure the purity of our dataset. On the other hand, the omission of these galaxies may restrict the domain in which the model can perform well.
We obtain a total of 66,092 postage-stamp images. Figure~\ref{fig:size} shows the distribution of the HST postage-stamp images prior to the $w=128$ cut. 

As mentioned above, our pre-training dataset is created by {\tt GalSim} using the characteristics derived from our HST dataset.
We choose to characterize HST galaxies by their best-fit elliptical Sersic parameters and fluxes.
The circular Sersic profile is defined as
\begin{equation}
    I(R) = I(R_{50}) \exp{\left(-b_n\left[\left(\frac{R}{R_{50}}\right)^{1/n}-1\right]\right)},
\end{equation}
where $n$ represents the Sersic index, $R$ is the radius, $R_{50}$ is the half-light radius, and $I(R_{50})$ is the surface brightness at $R_{50}$. 
The elliptical Sersic profile is implemented by replacing R with the following:  
\begin{equation}
\scriptstyle
    R = (ab)^{1/2}\left[\left(\frac{x}{a}\cos\theta + \frac{y}{a}\sin\theta\right)^2 +  \left(-\frac{x}{b}\sin\theta + \frac{y}{b}\cos\theta\right)^2\right]^{1/2},
\end{equation}
where $a$ and $b$ denote the semi-major and semi-minor axes, respectively.

Because of the rapid change in $I(R)$ near the center ($R=0$), evaluating $I(R)$ only at discrete positions (i.e., at the center of each pixel) leads to a significant discrepancy between the rendered and real galaxy images. To mitigate this, we first oversampled $I(R)$ with ten times higher resolution in the central region ($R<5$ pixel) and then applied $10\times10$ nearest neighbor-averaging to render the galaxy.
This Sersic galaxy image is convolved with the HST PSF and then fit to the HST galaxy image.
Fluxes are measured by summing up the background-subtracted pixel values from the HST galaxy images.

\subsection{GalSim Dataset} \label{subsec:Galsim}

For the generation of the pre-training dataset, 
we use {\tt GalSim} based on the distributions of the Sersic parameters derived from the HST dataset to
reproduce not only the marginalized distribution of each parameter but also the correlations between the parameters.
We constructed the probability density functions (PDFs) in the five dimensional parameter space (axis ratio, Sersic index, flux, size, and noise) with the 
{\tt scipy.stats} Gaussian kernel density estimator (KDE) module. Then, 100,000 new input parameter sets for {\tt GalSim} were generated by resampling based on the derived PDFs.

The image size $w$ varies in the original HST postage-stamp dataset. However, the galaxy images in the training dataset need to have an equal size. We choose a size of $64\times64$ pixels since the choice provides a good balance between efficiency and image fidelity. A smaller image would truncate the profile too prematurely while a larger image would be dominated by background values.

The 100,000 noiseless galaxy images output by {\tt GalSim} constitute the GT images in the pre-training dataset. 
Subsequently, these GT images were convolved with a 7$\times$7 pixel Gaussian kernel whose width matches the HST PSF.
Finally, Gaussian noise was added to align the rms noise level with that of the HST dataset. These degraded images comprise the LQ images.

\begin{figure}[ht!]
\plotone{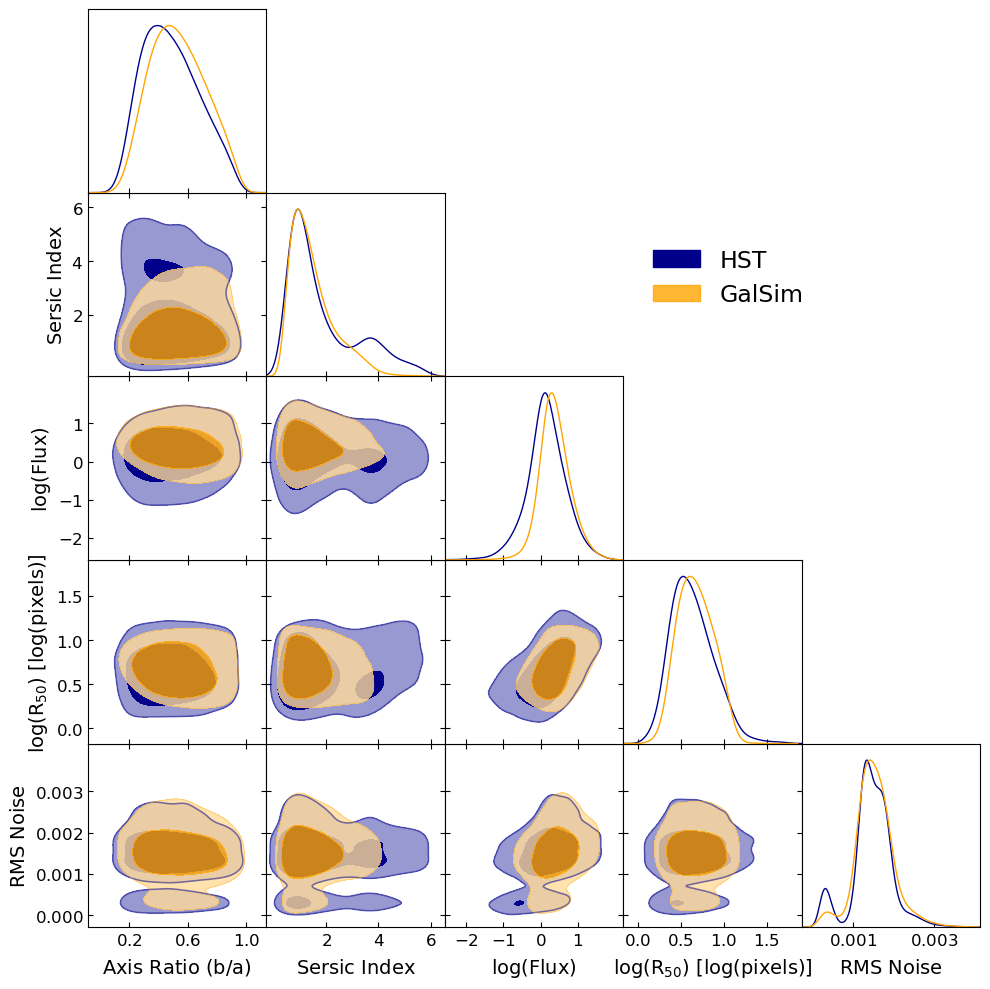}
\caption{Galaxy property comparison between the HST and {\tt GalSim} datasets.
After measuring the HST galaxy properties, we construct a 5-dimensional PDF. Input parameters for {\tt GalSim} are generated through resampling from the pdf while preserving correlations between parameters. Then, we determine the galaxy properties from the resulting {\tt GalSim} images with the same procedure used in the HST case. 
Overall, the {\tt GalSim} galaxy images follow the HST properties reasonably well. However, the {\tt GalSim} distribution is inevitably smoother because
since the input parameters as well as the {\tt GalSim} images contain noise.}
\label{fig:corner}
\end{figure}

Following the same procedure used for determining the best-fit Sersic parameters from the HST galaxy images, we measured the Sersic parameters from the LQ images.
Figure~\ref{fig:corner} compares the Sersic parameter distributions between the HST and {\tt GalSim} LQ images.
One must understand that in principle the two distributions cannot align exactly because the input Sersic parameters used for the {\tt GalSim} image generation are derived from the noisy HST images.
Nevertheless, we find that overall the covariances and marginalized probability distributions are similar between the two datasets. Since the {\tt GalSim} galaxy images are used to pre-train the model, which is subsequently finetuned by the JWST-based images, we do not think that the differences in detail matter.

\subsection{JWST Dataset} \label{subsec:JWST}
We applied nearly the same procedures used for creating the HST dataset to sample the JWST image. The only difference is the criteria for the noise level. Since the JWST images prior to degrading serve as GT, the noise level must be considerably lower. We accepted an image only if its rms noise after the min-max normalization became less than 0.02.

One non-trivial issue is how to handle the difference in pixel scale between JWST and HST.
NIRCAM's native pixel scale is  $0.03\arcsec$ ($0.06\arcsec$)
for the short (long) wavelength channel, which is different from the HST/ACS pixel scale $0.05\arcsec$.
We considered resampling the JWST pixel scale to $0.05\arcsec$. However, the interpolation noise degraded the original quality non-negligibly. Therefore, we maintained the original scale. Therefore, our JWST galaxies are 40\% larger or 20\% smaller than the HST ones on average.
However, as we demonstrate in \textsection\ref{sec:results}, our HST image restoration is not significantly affected by this pixel scale inconsistency.

Another tricky issue is the JWST's intrinsic PSF. Although JWST's PSF is on average much smaller than that of HST, its long wavelength channel PSF width can be similar to that of the HST/ACS PSF. One may consider deconvolving JWST image with its own PSF as an attempt to further enhance its resolution and thus better represent the truth.
Perhaps this is achievable with an approach similar to the current method explored here. However, this is beyond the scope of this study. In the current investigation, we use JWST images without deconvolution. Thus, the effective PSF of the LQ images that we describe below includes the convolutions by both JWST's and HST's PSFs.

We standardize the postage-stamp image size to $64\times64$ pixels. A smaller image is expanded by padding pixels to its four edges in such a way that both the noise and background levels match the original image. A larger image is trimmed to $64\times64$ pixels. We discarded the galaxy images whose profiles were prematurely truncated by the trimming procedure.
We collected 113,485 postage-stamp galaxy images from the JWST NIRCam F115W, F200W, F277W, and F444W images from various fields publicly available in August 2023 the Mikulski Archive for
Space Telescopes (MAST)\footnote{\url{https://archive.stsci.edu/}}: \href{https://mast.stsci.edu/portal/Mashup/Clients/Mast/Portal.html?searchQuery=%7B%22service%22:%22DOIOBS%22,%22inputText%22:%2210.17909/crbp-3069%22%7D}{ 10.17909/crbp-3069}. The dataset serves as the GT component in our finetuning dataset.

The LQ counterpart is generated by convolving the JWST GT galaxy images with the HST PSF and adding Gaussian noises. Although the pixel scale is different from the HST image, we scaled the HST PSF size, assuming the $0.05\arcsec$ pixel scale. This is because, eventually, we desire to restore the HST images whose pixel scale is $0.05\arcsec$. The applied noise distribution follows that of the HST dataset.
The ratios among the training, validation, and test datasets are 8:1:1.
The GalSim and JWST train set subsamples are publically available.\footnote{\url{https://drive.google.com/file/d/1dOemrQXFr2UxHgvUTUffLqIXlHeEkRbZ/view?usp=sharing}}

\section{JWST Test Dataset Results} \label{sec:results}
In this section, we present the results when our deep learning model is applied to the JWST test dataset. In this case, for each LQ image, there exists a corresponding GT image, which makes the comparison straightforward. However, readers are reminded that when we apply our model to real images (\textsection\ref{sec:realapplication}), where there are no matching GT images, we can only use GT proxies obtained also from real images for comparisons.

Below, we define the two evaluation metrics: peak signal-to-noise ratio (PSNR) and structural similarity index measure (SSIM) 
(\textsection\ref{subsec:psnr}), 
describe our visual inspection 
(\textsection\ref{subsec:visual}), and compare morphological (\textsection\ref{subsec:morphology}) and photometric properties (\textsection\ref{subsec:photometry}).

\subsection{PSNR and SSIM\label{subsec:psnr}}
The PSNR and SSIM metrics are employed to evaluate the similarity between LQ and GT images and guide the selection of our best models. The mean squared error (MSE) is defined as:
\begin{equation}
    \text{MSE} = \frac{1}{mn}\sum_{i=1}^{m}\sum_{j=1}^{n} [x(i,j) - y(i,j)]^2,
\end{equation}
where $m$ and $n$ represent the size of the image, $x(i, j)$ [$y(i, j)$] is the pixel value at the position $(i,j)$ in the GT (RS) image.
MSE measures the average squared difference between corresponding pixels in two images. 
With the MSE defined above, PSNR is given by:
\begin{equation}
    \text{PSNR} = 10 \cdot \log_{10}\left(\frac{\mbox{MAX}^2}{\text{MSE}}\right),
\end{equation}
where MAX is the peak value of the GT image.
Compared to MSE, PSNR takes into account the image-by-image difference in peak values.

The SSIM metric is defined as:
\begin{equation}
    \text{SSIM}(x, y) = \frac{(2 \mu_x \mu_y + C_1)(2 \sigma_{xy} + C_2)}{(\mu_x^2 + \mu_y^2 + C_1)(\sigma_x^2 + \sigma_y^2 + C_2)},
\end{equation}
where $\mu_x$ ($\mu_y$) and $\sigma_x^2$ ($\sigma_y^2$) are the mean and variance, respectively, of the $x$ ($y$) image, and $\sigma_{xy}^2$ is the covariance between the $x$ and $y$ images.
$C_1$ and $C_2$ are introduced to prevent the equation from diverging when the denominator is small, and the exact values depend on the dynamic range of the target images.
SSIM is a perceptual metric that evaluates the structural similarity between two images and considers luminance, contrast, and structure. The value ranges from -1 to 1 with 1 being the perfect similarity.

In Table~\ref{table:comparisonTable}, we compare the PSNR and SSIM values between our fiducial model (transfer learning) and experimental (non-transfer learning) models, including the values measured directly between GT and LQ datasets without model training. The transfer learning model provides the highest values in both PSNR and SSIM.
Although in terms of PSNR the transfer learning shows only a marginal ($\mytilde0.33$\%) improvement over the JWST-only training case, the improvement in SSIM is
4.67\% when normalized by the SSIM value for the no-training case.

\begin{deluxetable}{ccC}
\tablenum{1}
\tablecaption{Comparison of PSNR and SSIM performance between the model trained through transfer learning and the models trained with either the {\tt GalSim} or JWST dataset.\label{table:comparisonTable}}
\tablehead{
\colhead{Training Dataset} & \colhead{PSNR} & \colhead{SSIM}
}
\decimalcolnumbers
\startdata
GalSim &  32.2997 & 0.4711 \\
JWST & 38.7868 & 0.8534 \\
\textbf{GalSim $\rightarrow$ JWST} &  \textbf{38.8416} & \textbf{0.8622} \\
no training & 16.4488 & 0.1883 \\
\enddata
\tablecomments{The model obtained with transfer learning (bold) shows improvements in both PSNR and SSIM. The last row presents the PSNR and SSIM measured directly between the GT and LQ datasets, displaying the initial quality of the datasets.}
\end{deluxetable}

\subsection{Visual Inspection\label{subsec:visual}}
\begin{figure}[!htb]
\includegraphics[width=1\linewidth]{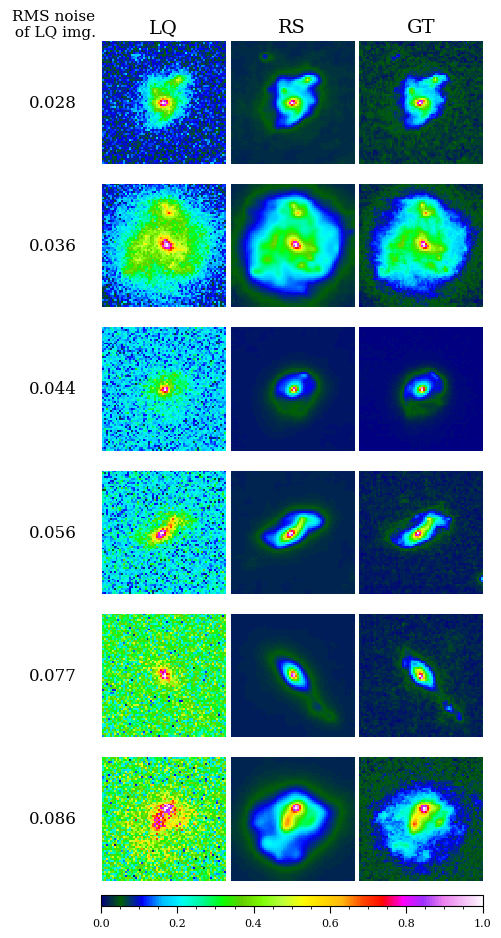}
\caption{Examples of restored images. We compare the low-quality (left), restored (middle), and ground-truth (right) images. 
Here, we select examples that have rich substructures, low-surface brightness features, and extended morphologies.
From top to bottom, the RMS noise level increases.
Each galaxy's overall shape, low-surface brightness features, and substructures are remarkably well restored regardless of the noise level in the low-quality image.}
\label{fig:visual}
\end{figure}

Figure \ref{fig:visual} shows several image restoration cases in ascending order of rms noise level. We select examples, which have rich substructures, low-surface brightness features,
and extended morphologies. 
The visual inspection indicates remarkable improvements in both resolution and noise level.
Detailed substructures such as star-forming clumps and spiral arms have been restored with high fidelity. Also, the restored overall morphological features such ellipticity and core size are consistent with those of the GT images. 
In addition, the low-surface brightness edges are restored remarkably well.
Finally, we note that the performance is not very sensitive to the input noise level, which in this example varies by a factor of 3.

\begin{figure*}[ht!]
\centering
\includegraphics[width=1\linewidth]{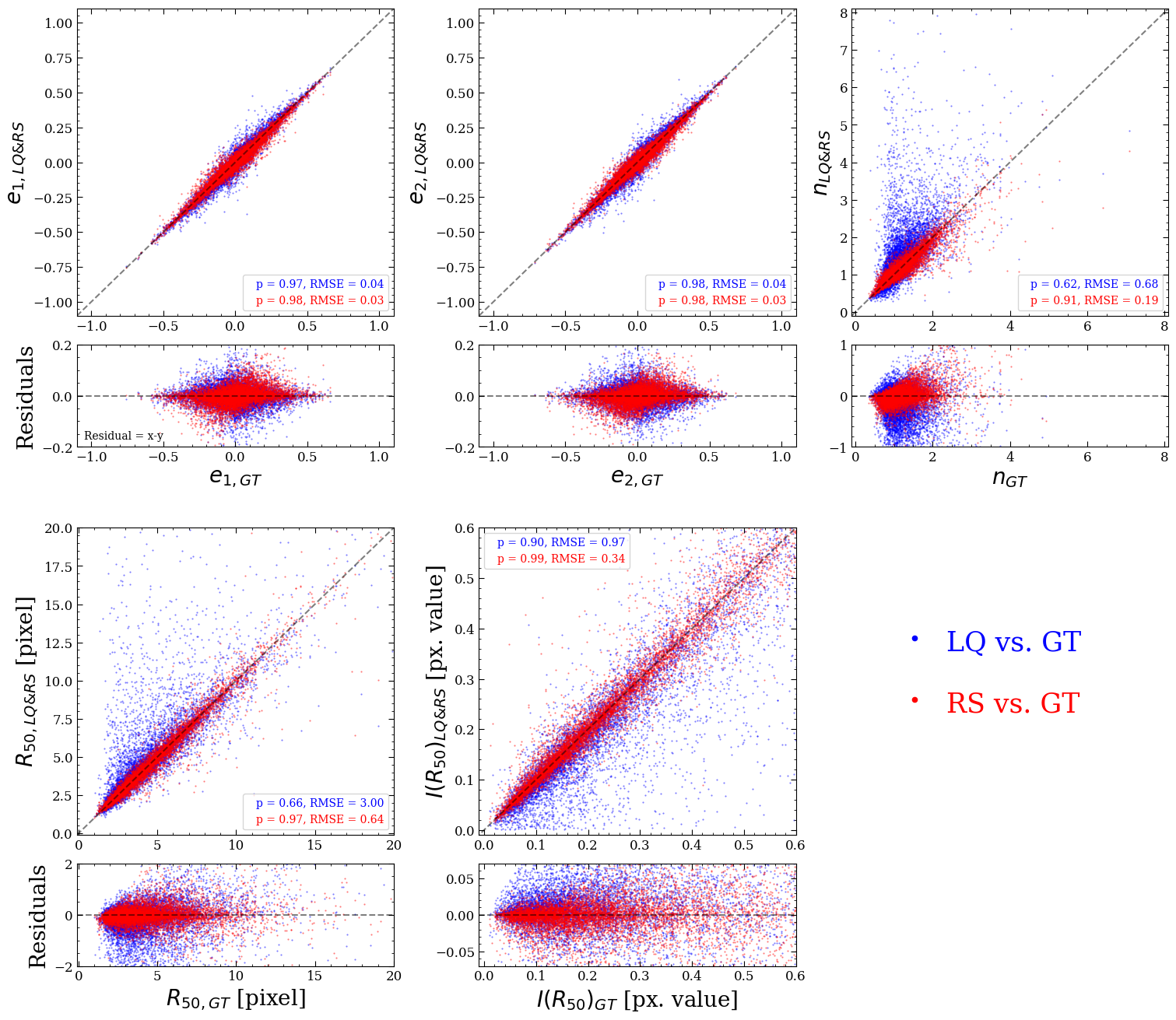}
\caption{Comparison of morphological parameters between GT and LQ images (blue) and between GT and RS images (red).
We investigate $e_1$, $e_2$, $n$ (Sersic index), $R_{50}$ (half-light radius), and $I_{R50}$ (intensity at $R_{50}$) determined from the Sersic fitting.
Black dashed lines denote one-to-one correlations.
The RS images have a stronger correlation ($p$) with a smaller scatter (RMSE).}
\label{fig:morphology}
\end{figure*}

\subsection{Restoration of Morphological Parameters\label{subsec:morphology}}
Although one can characterize galaxy morphologies in various ways, we employ best-fit Sersic parameters to quantify the comparison. Specifically, we compare the two ellipticity components: $e_1$ and $e_2$, Sersic index $n$, half-light radius $R_{50}$, and the intensity at $R_{50}$.

The two components of the ellipticity are motivated by the weak lensing convention, which utilizes not only the absolute ellipticity $e$ of a galaxy but also its position angle $\theta$ as follows:
\begin{equation}
    e_1 = e\cos(2\theta),
\end{equation}
\begin{equation}
    e_2 = e\sin(2\theta).
\end{equation}
The ellipticity $e$ is defined as $(a-b)/(a+b)$, where $a$ and $b$ are the semi-major and -minor axes, respectively.

Figure \ref{fig:morphology} compares the five morphological parameters.
Across all parameters, the RS images exhibit stronger correlations with the GT images in terms of the scatter and slope of the correlations.
The ellipticity comparison shows that the LQ galaxies are systematically rounder than the GT galaxies, which is not surprising because they are generated by convolving the GT galaxies with the circular HST PSF. This bias nearly disappears in the RS images.
The improvement in the Sersic index $n$ is remarkable. The scatter is reduced by more than a factor of 3. Also, the Pearson correlation coefficient improves from  0.61 to 0.90. Given that the Sersic index $n$ is one of the most difficult parameters to restore, its recovery showcases the stability of the restoration performance. The $R_{50}$ recovery is also noteworthy with the reduction of the scatter by more than a factor of 4 and the increase of the Pearson correlation coefficient by $\mytilde48$\%. Finally, the $I(R_{50})$ intensity comparison shows a scatter reduction of $\mytilde46$\% and an increase of the correlation by $\mytilde9$\%.

\begin{figure*}[ht!]
\centering
\includegraphics[width=1\linewidth]{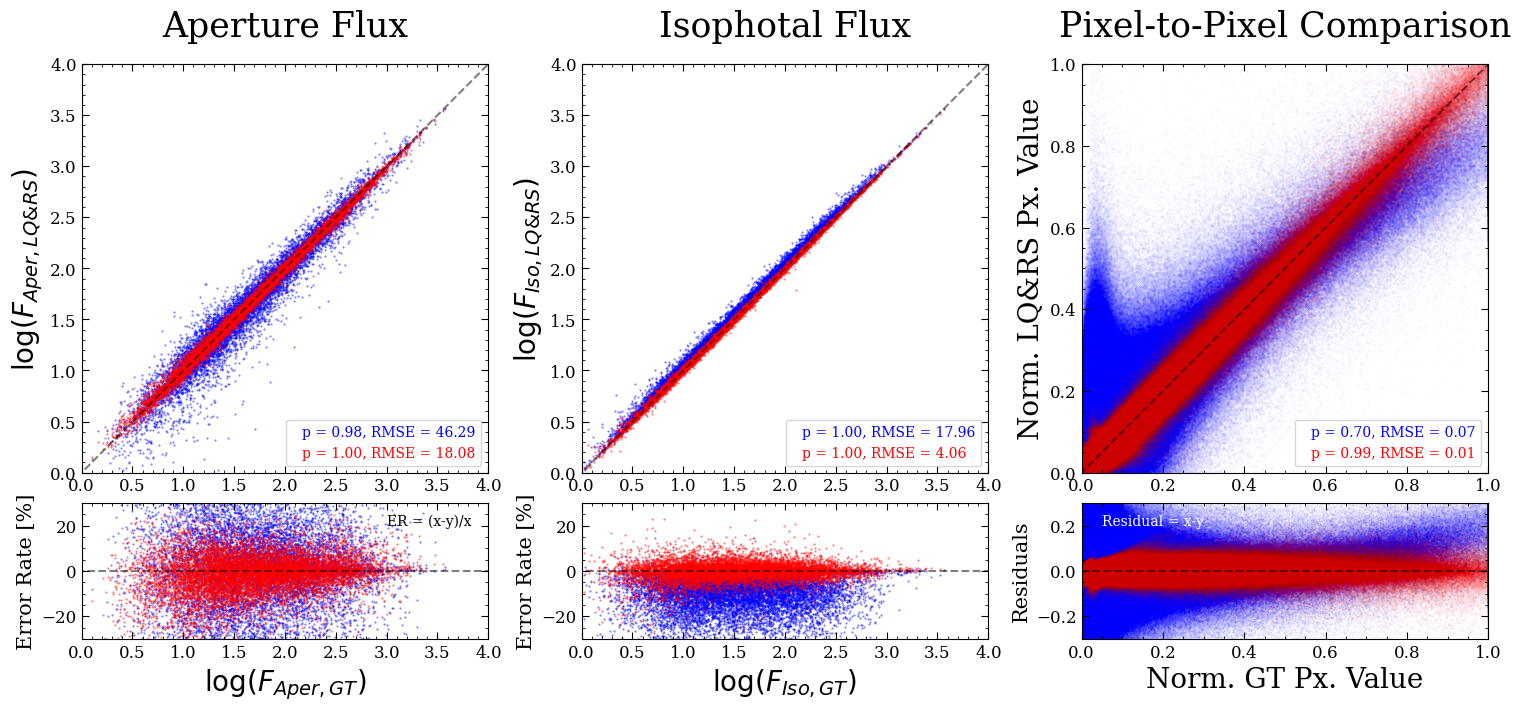}
\caption{Comparison of photometric information between GT and LQ images (blue) and between GT and RS images (red).
We investigate correlations of aperture flux, isophotal flux, and individual pixel values. 
For flux comparison, since the RS images were scaled to the range $[0,1]$, we rescaled the RS images to enable quantitative comparisons (see text for details). We stress that we {\it do not} use the information from GT for rescaling.
We use an elliptical aperture defined with {\tt SExtractor}'s semi-major, semi-minor axes, and orientation angle from the LQ image. The isophotal area is also determined from the LQ image.
For comparison of individual pixels, we use the pixels only within the elliptical aperture. Flattened pixel values in LQ images due to convolution are restored to their original values, reproducing a one-to-one slope.
The photometric information is recovered remarkably well, with a significant reduction of scatter compared to the LQ-GT comparison.}
\label{fig:photometry}
\end{figure*}

\subsection{Restoration of Photometric Parameters\label{subsec:photometry}}
One of the immediate scientific utilities of image restoration is enhancing photometry.
Here we compare aperture flux, isophotal flux, and individual pixel values measured from the RS images with the GT images to evaluate the performance in the photometric context.

Since we use min-max normalization consistently across our training and input datasets, the dynamic range of the RS images is also restricted. To extract photometry from the LQ and GT images, we opt to use the original (pre-normalization) images. Consequently, it is necessary to rescale the RS images. To ensure a fair comparison, this rescaling process must be executed independently, without relying on the information available from the corresponding GT images.

We aligned the dynamic range of the RS images to the LQ images as follows.
First, we measured the lower (background) level $l_{LQ}$ from the LQ image using the combination of $\sigma$-clipping and mode estimation implemented in {\tt SExtractor}.
The upper level of the LQ image $u_{LQ}$ is determined by averaging the central $4\times4$ pixel values. Then, we convolved the RS image with the HST PSF to match the LQ image resolution. With the same procedure, we measured the lower and upper levels ($l_{RS}$ and $u_{RS}$, respectively) from this PSF-convolved RS image.
Finally, we rescaled the RS image by first multiplying $s$ and then adding $c$, where
\begin{equation}
    s=\frac{u_{LQ}-l_{LQ}}{u_{RS}-l_{RS}}
\end{equation}
and
\begin{equation}
    c=-sl_{RS} +l_{LQ}.
\end{equation}

For aperture flux measurement, We defined a Kron ellipse using the LQ image using {\tt SExtractor} and applied it consistently across the LQ, RS, and GT images.
Similarly, isophotal areas were defined from the LQ image and used consistently across the LQ, RS, and GT images. 

Figure~\ref{fig:photometry} displays the resulting flux comparisons. The aperture fluxes from the RS images are in good agreement with those from the GT images. Compared to the LQ images, the scatter is reduced by $\mytilde60$\%. The scatter reduction is similar ($\mytilde55$\%) in isophotal flux. It is worth noting that the isophotal fluxes in the LQ images are systematically overestimated because the isophotal 
area is defined from the LQ image\footnote{That is, noise can make some pixel values near the edge of the isophotal area in the LQ image higher than the GT values.}.
This bias is significantly reduced in the RS images.
Finally, the pixel-to-pixel comparison illustrates a tight 1:1 correlation between the RS and GT images across the entire dynamic range, while the LQ images show a slope significantly less than unity because of their larger PSF. The pixel-to-pixel scatter reduction is by a factor of 7. 

\section{APPLICATION TO REAL HST IMAGES} \label{sec:realapplication}

\begin{figure*}[t!]
\includegraphics[width=1\linewidth]{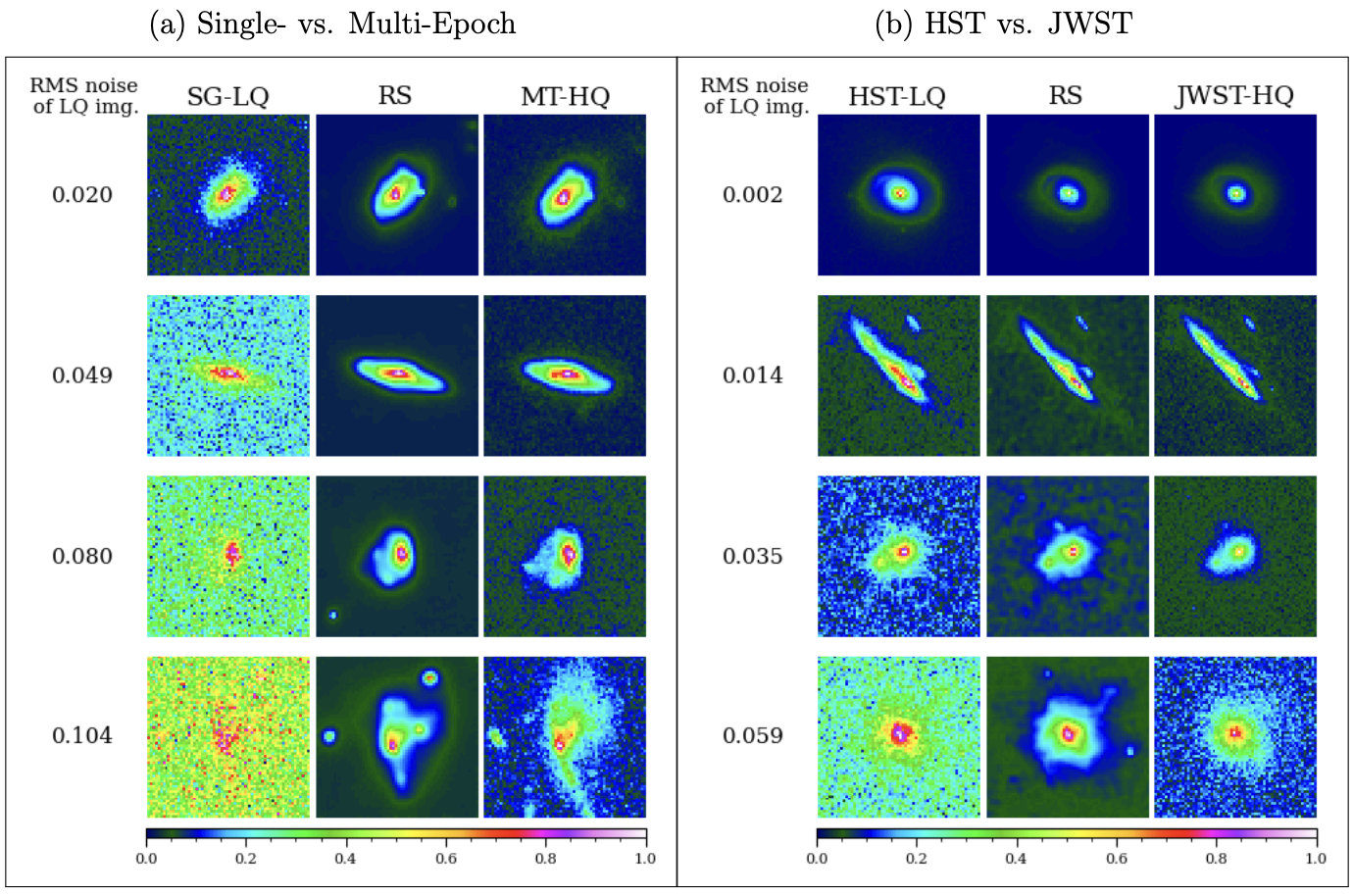}
\caption{Application to HST images. We apply our restoration model discussed in \textsection\ref{sec:Method} without any modification to HST images.
(a) Comparison of single-epoch RS images with multi-epoch high-quality images. Here, we use deep multi-epoch images as a proxy for GT. The single-epoch LQ images were enhanced using our model. Note that since the model is designed to perform deconvolution to match the JWST PSF, the resolution of the RS images is higher than that of the deep multi-epoch HST images. 
Also, the noise levels in the RS images are lower than in the multi-epoch data. 
(b) Comparison of multi-epoch RS images with deep JWST images. Here, we use deep JWST images as a proxy for GT. Since the JWST/NIRCam images are Near IR (F115W) images while the HST/ACS images are optical (F606W) images,
only limited comparison is possible.
Nevertheless, both the reduction in noise and improvement in resolution make the RS images look more similar to the JWST images than the corresponding LQ images.}
\label{fig:HST inference}
\end{figure*}

In \textsection\ref{sec:results}, we have demonstrated that our deep learning model successfully restores the simulated HST images to the JWST-quality level in both resolution and depth. Here we present our results when the same model is applied to real HST images. It is important to note that we cannot perform identical comparisons as presented in \textsection\ref{sec:results} because there are no equivalent GT images for the real HST images. In \textsection\ref{subsec:single}, we restore single-epoch (short-exposure) images and compare their RS images to multi-epoch (long-exposure) images, using the latter as proxies for the GT dataset.
Since our deep learning model not only reduces the noise level but also the effective PSF, the comparison cannot be fair because the multi-epoch image suffer from the same HST PSF. In addition, the multi-epoch image is not sufficiently deep and often noisier than the RS images.
\textsection\ref{subsec:single}, we restore HST images and compare their RS images to deep JWST images. In this case, the resolutions between the two datasets are aligned. However, since their filters are different, we cannot expect the pairs to have identical photometric and morphological properties even if the deep learning model provides perfect restoration.

\subsection {Enhancement of Single-epoch Images and Comparison with Multi-epoch Images\label{subsec:single}}

We utilize the datasets from the HUDF and  GOODS-N$\&$S observed by the HST ACS.
The single-epoch low-quality (SG-LQ) images are retrieved from the MAST archive. We aim to enhance the {\tt FLT} images not their ``drizzled" ({\tt DRZ})images because ``drizzling" induces inter-pixel noise correlations. When a postage-stamp
image is contaminated by cosmic rays, we remove them with the {\tt  lacosmic} script\footnote{https://lacosmic.readthedocs.io/en/stable} \citep{vandokkum2001, lacosmic}.
Since the multi-epoch high-quality (MT-HQ) mosaic images are corrected for geometric distortions unlike the {\tt FLT} images, we transformed them back to the distorted, CCD coordinate system to enable a fair comparison with the {\tt FLT} images.

In Figure \ref{fig:HST inference}a, we display four enhancement examples. Similar to the results observed in the experiment with the JWST test dataset, the restoration qualities are similarly excellent here. The low-surface brightness features, which cannot be identified in the LQ images, are well-restored in the RS images. One notable difference from the results in the JWST test dataset is the PSF size. Since our deep learning model is trained to restore the JWST resolution, the PSF size of the RS images are smaller than the HQ images. Another important difference is on average the higher noise level in our single-epoch LQ images.

\begin{figure*}[p!]
\includegraphics[width=1\linewidth]{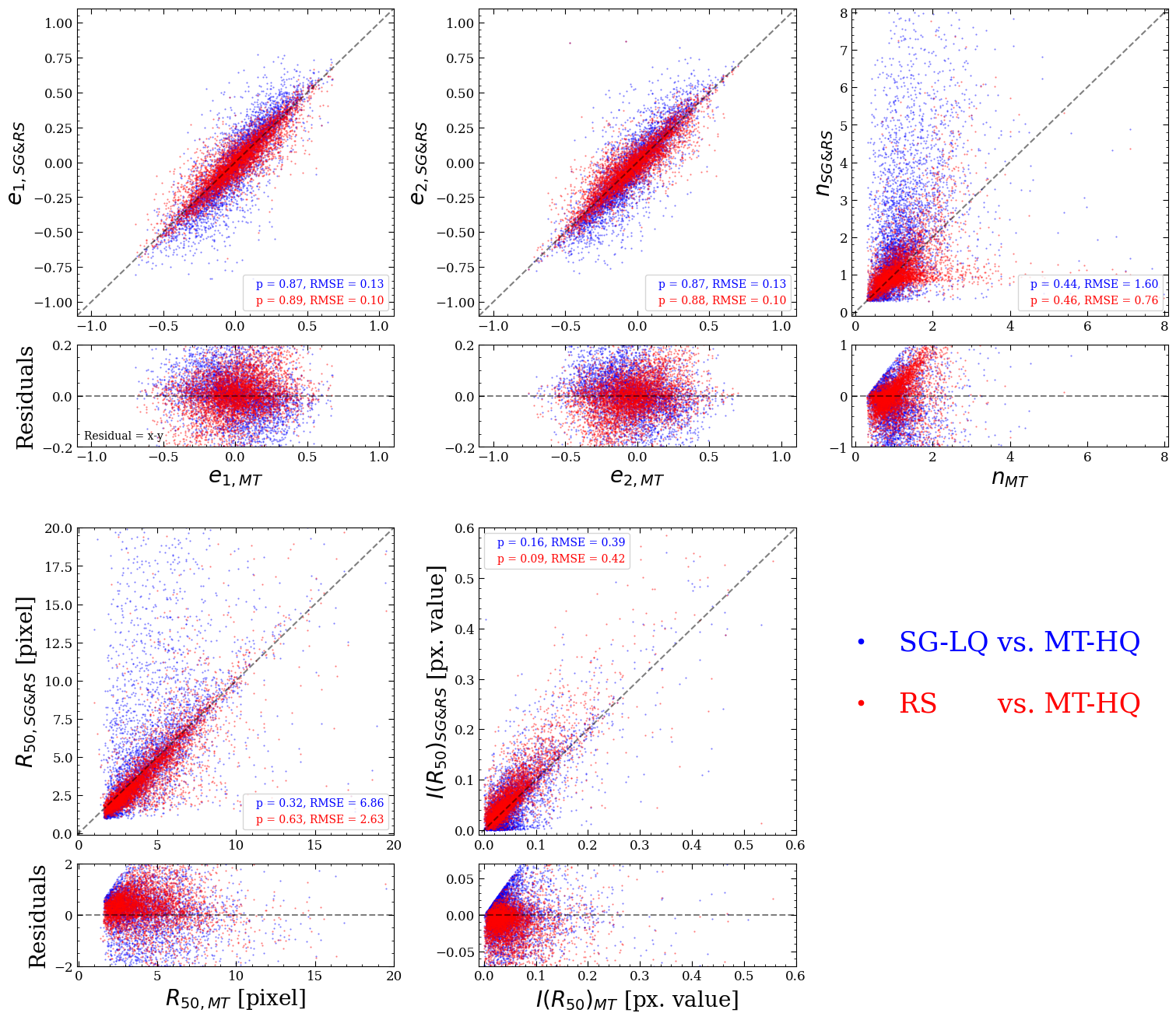}
\caption{Comparison of morphological parameters between MT-HQ and SG-LQ images (blue) and between MT-HQ and RS images (red). We follow the same format as in Figure \ref{fig:morphology}. Note that the MT-HQ images have higher noise than the GT images created from the JWST images. 
Also, the SG-LQ images are noisier than the LQ images in the JWST dataset.
Since SG images are in the geometrically distorted CCD coordinate, we applied the due transformation
to the MT-HQ images in the rectified coordinate to align them with the SG images in the CCD coordinate.
The properties measured in RS images are more closely correlated with those in MT-HQ images than those in the SG-LQ images.
}
\label{fig:hst_morphology}
\end{figure*}

Figures \ref{fig:hst_morphology} shows a comparison of the five morphological parameters. Significant improvements over the LQ images are clear in terms of the correlation strength and reduction of the scatter. Readers are reminded that we used the multi-epoch image as a proxy for the GT image after applying an inverse geometric distortion correction to it. Therefore, compared to the JWST dataset experiment, the measurement of the five morphological parameters from the GT surrogate is substantially noisier here. Also, as mentioned above, the PSF size of the RS image is smaller than that of the HQ image, which can contribute to a systematic difference in shapes between the two datasets.

We display a comparison of photometric parameters in Figure~\ref{fig:photometry}. Again, significant improvements over the single-epoch LQ images are apparent. In particular, the pixel-to-pixel comparison shows a remarkable difference in both correlation strength and scatter reduction. Since the caveats mentioned in the morphological parameter comparison above apply here too, the improvements are less pronounced than those in the experiment with the JWST test dataset.

\subsection{Enhancement of Multi-epoch HST Images and Comparison with Multi-epoch JWST Images\label{subsec:hst_vs_jwst}}

We utilize the publicly available HST and JWST multi-epoch mosaic images of Abell 2744. Specifically, we apply our Transformer model to the HST ACS/WFC F606W images and compare their RS images with the JWST NIRCam F115W images.
We resampled the JWST images to match the HST pixel scale (0.05$''pix^{-1}$).
Since galaxies must have different morphological and photometric parameters between optical and near IR, we provide only a visual comparison here.

Figure~\ref{fig:HST inference}b shows four examples. Despite the difference in filter, the RS images look remarkably similar to the JWST images. In particular, when the noise level is relatively low (i.e., top two panels), the enhancement in both resolution and low-surface brightness feature is excellent. When the noise level is relatively high, the RS images still resemble the JWST images more closely than their input HST images. We suspect that the correlated noise in the input multi-epoch HST image might have non-negligible impact on the result in this noise regime.

\begin{figure*}[ht!]
\includegraphics[width=1\linewidth]{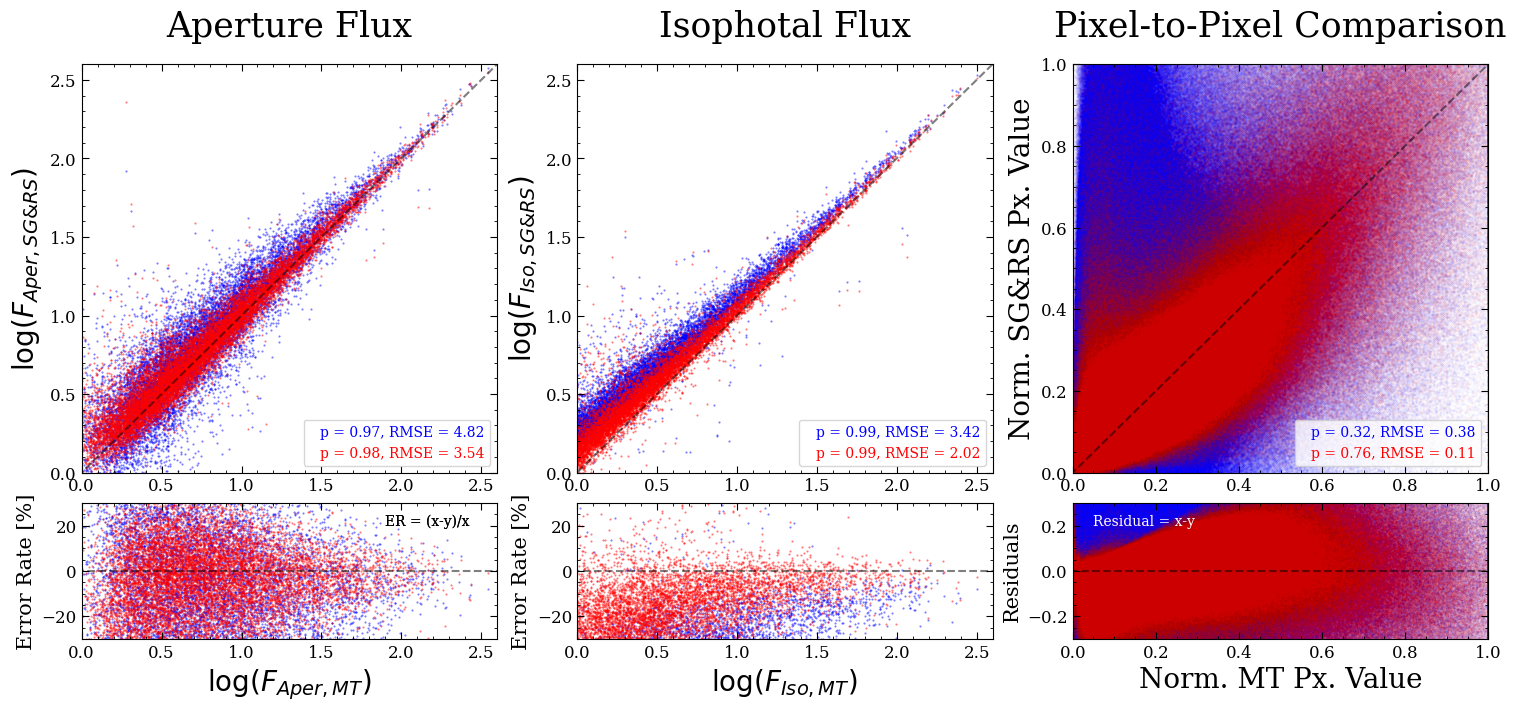}
\caption{Comparison of photometric information between MT-HQ and SG-LQ images (blue) and between MT-HQ and RS images (red). 
Significant improvements over the SG-LQ images are clear. In particular, the
pixel-to-pixel comparison shows remarkable enhancement
in both correlation strength and scatter reduction.
}
\label{fig:hst_photometry}
\end{figure*}

\section{LIMITATIONS\label{sec:limitation}}
\subsection{Degradation in Restoration Quality Due to High Noise Level\label{subsec:fail}}
Although our Transformer-based deep learning model provides state-of-the-art performance in both resolution enhancement and noise reduction for moderate noise levels, inevitably, proper restoration becomes impossible when the noise level exceeds a certain threshold.

\begin{figure}[htb!]
\includegraphics[width=1\linewidth]{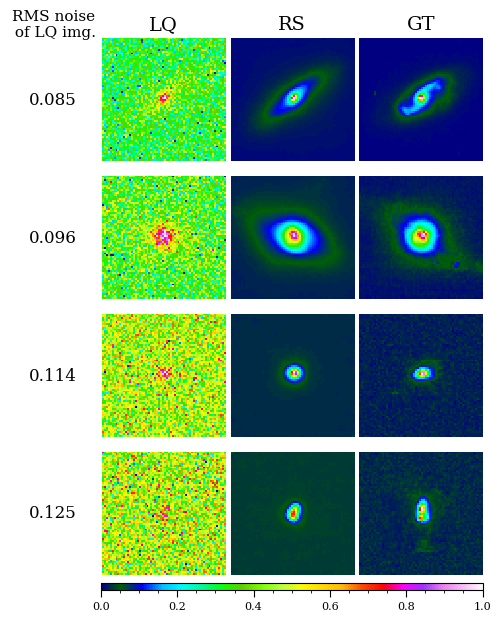}
\caption{Performance degradation cases.
As the noise level increases,
the restoration depends on fewer and fewer pixels.
Thus, more frequent degradation cases occur with higher noise levels. Shown here are examples of loss of spiral arms (first row), incorrect ellipticity (second and third row), and failed de-blending of multiple peaks (fourth row). 
}
\label{fig:failures}
\end{figure}

Figure~\ref{fig:failures} shows four such examples. The top row illustrates a case, where the distinct spiral arm in the GT image is not restored. Upon scrutiny of the corresponding LQ image, we believe that 
the information loss due to noise is too significant to hint at the presence of the spiral arm.
The second row shows a case, where the position angle of the GT image is not properly restored. Again, we suspect that the LQ image is too noisy to enable a proper inference of the position angle of the GT image.
The third row is an example, where the ellipticity of the RS image is much rounder than that of the GT image. In the last row, we illustrate a case, where the deep learning model fails to unveil the two distinct peaks of the GT image.

Since the exact definition of restoration failure is a subjective matter, it is difficult to quote an exact threshold. Nevertheless, from visual inspections, we suggest that the failure frequency increases noticeably when the rms value (after min-max normalization) of the LQ image is larger than $\mytilde0.1$.

\subsection{Point Source Recovery Test\label{subsec:point_test}}
In principle, the perfect deconvolution algorithm should restore a point source to a delta function, which is infinitely smaller than a pixel. In traditional deconvolution in the Fourier domain, it is a challenging task because the operation is numerically unstable. The resulting images often exhibit many ringing effects around a bright central peak. 

Since we excluded stars from the training dataset, our deep learning model did not explicitly learn to deconvolve point source images. Thus, it is interesting to examine how well our deep learning model, trained with only galaxy images, restore point sources.
We perform a point source recovery test as follows. First, we created 1,000 JWST-quality star GT images. Because we did not remove the JWST PSF from the training dataset, the star GT images should not resemble a delta functions but the JWST PSF. We implemented this by convolving a single pixel with a Gaussian whose kernel size matches the JWST PSF. Note that we randomize the positions of the stars within the central 24$\times$24 pixels of the $64\times64$ postage-stamp images. Then, we created their LQ versions by further convolving the GT images with the HST PSF and adding noise. Finally, these LQ images are restored by our deep learning model.
To investigate the systematic effect, we stack the 1,000 GT and RS images separately after aligning their centers. 

\begin{figure}[htb!]
\includegraphics[width=1\linewidth]{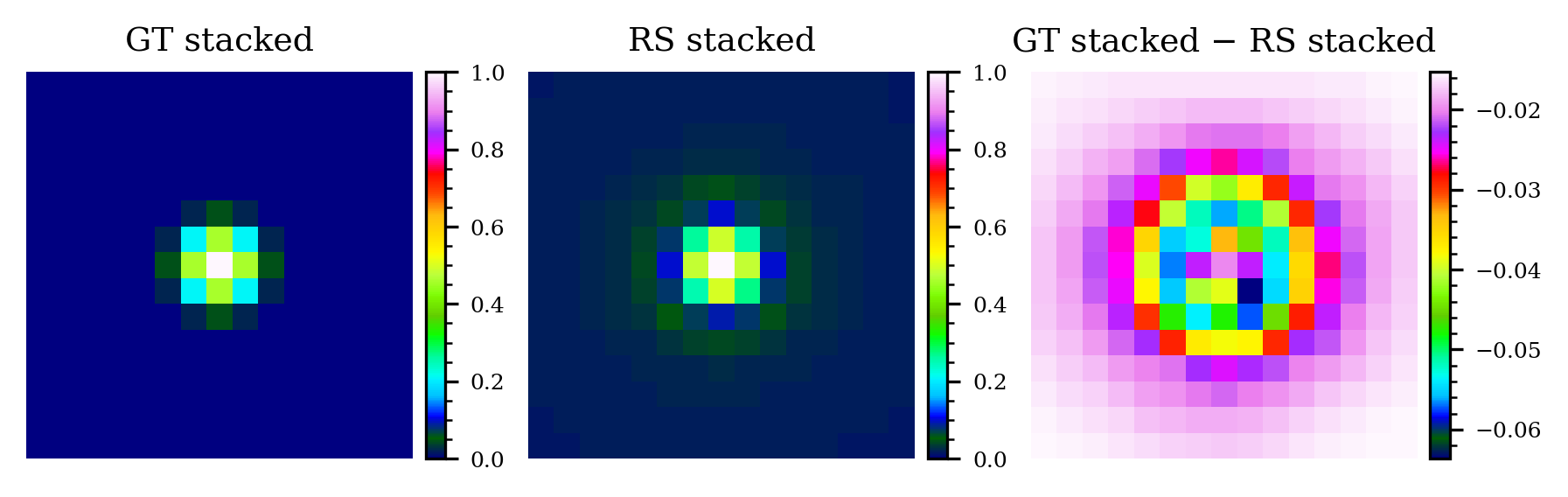}
\caption{Point source restoration test.
We created 1000 LQ-GT pairs of point source images and compared the mean RS image with the mean GT image (see text for details).
The profile in the RS image is slightly more extended than that in the GT image.}
\label{fig:point}
\end{figure}

Figure~\ref{fig:point} shows that the difference is subtle when we compare the GT (left) and RS (middle) stacks visually. However, the residual image (right) illustrates that the PSF of the RS stack is systematically larger. Thus, we conclude that our deep learning model, trained solely with galaxy images, performs less than ideally for point sources.

\subsection{Artifacts Due to Pixel Correlation
\label{subsec:pixel_correlation}}
In our generation of LQ images, we assume that the noise is Gaussian. However, in real astronomical images, especially when we create deep images by stacking many dithered exposures, there exist significant inter-pixel noise correlations. We find that these inter-pixel noise correlations create non-negligible artifacts.

\begin{figure}[htb!]
\includegraphics[width=1\linewidth]{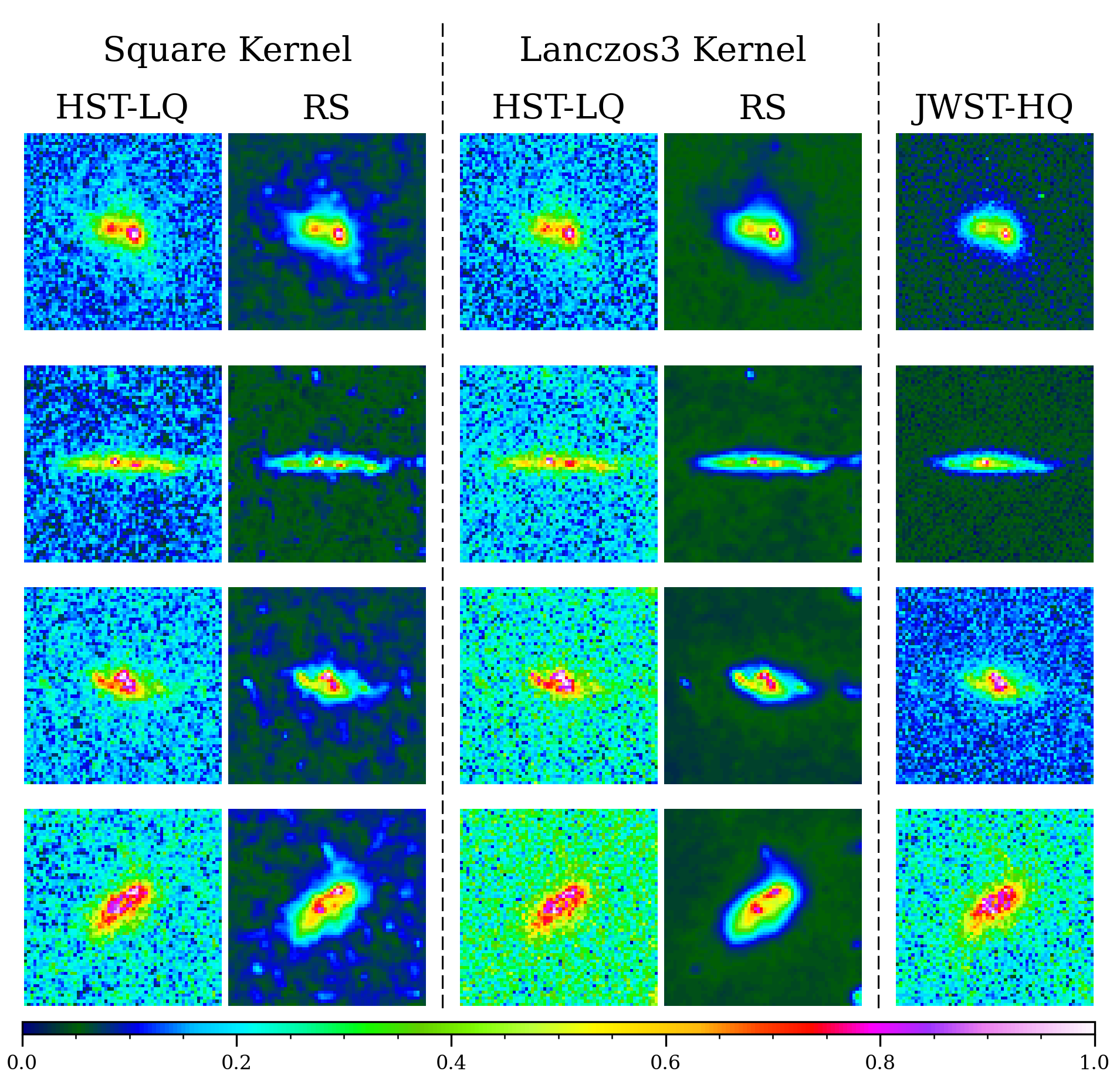}
\caption{Application to HST galaxies sampled from multi-epoch drizzled images. The first and third columns are the HST-LQ multi-epoch images drizzled with square and Lanczos3 kernels, respectively. Pixel values are significantly correlated for the LQ image produced with the square kernel. In their corresponding RS images, these correlations manifest as numerous artifacts, which are absent in the JWST-HQ images. The Lanczos3 kernel reduces pixel correlations in the LQ images and also minimizes artifacts in the corresponding RS images.}
\label{fig:noise_corr}
\end{figure}

Figure~\ref{fig:noise_corr} displays some examples of these artifacts. The LQ images here are sampled from multi-epoch images drizzled with square and Lanczos3 kernels using {\tt AstroDrizzle}~\citep{fruchter2010} in {\tt DrizzlePac}~\citep{hoffmann2021}. With square-kernel drizzling, correlated noise is apparent even from visual inspection. The RS images show that the correlated noise creates some low-surface brightness artifacts in the galaxy outskirts, which are absent in the JWST images. These artifacts are mitigated when we use the Lanczos3 kernel, which helps reduce the pixel correlation in the mosaic-drizzled image. The background and edge of the galaxy in the RS images are clean and smooth. Addressing this issue could involve strategies such as employing a different drizzling kernel for image stacking or leveraging more advanced deep learning algorithms. Exploring these solutions will be a key focus of our future work.

\subsection{Discussion on Potential Hallucination Effects
\label{subsec:hallcination}}
A hallucination effect in generative models refers to the phenomenon where the model generates outputs that are not grounded in reality or the input data, but instead are fabricated or ``hallucinated" by the model, leading to false interpretations. Since virtually all generative models are subject to hallucination effects in varying degrees, we discuss potential hallucination effects in our restoration model.

A detailed examination of Figure \ref{fig:HST inference}a reveals that the restored images of the single-epoch data in the third and fourth rows show some compact, point-like sources. Because these features do not exist in the corresponding much deeper multi-epoch data, they may be regarded as arising from hallucination effects. However, scrutiny of the corresponding LQ images indicates that the pixel levels of the regions where the hallucination effect is suspected are elevated compared to the neighboring regions. They may be outliers in noise fluctuations, uncorrected bad pixels, or residual cosmic rays that were not identified. We verified that these compact, point-like features do not appear in the RS images without their counterparts in the GT images when we restored images from the LQ images generated by degrading the JWST GT images (Figure~\ref{fig:visual}).
Therefore, we conclude that the artifacts in Figure~\ref{fig:HST inference}a are grounded in the real features of the original LQ images rather than caused by the hallucination effect.
Nevertheless, the discussion provides an important lesson: when we apply our model to real LQ images, we need to ensure that the input images do not possess glitches (e.g, bad pixels, cosmic rays, etc.) distinct from general noise.

\section{Conclusions}\label{sec:conclusion}
We have showcased astronomical image restoration from HST quality to JWST quality
using the efficient Transformer model via transfer learning. The pretraining dataset was created by rendering GT galaxy images based on analytic profiles and generating corresponding LQ versions by reducing their resolution and introducing noise.
The finetuning dataset was produced by sampling GT galaxy images from deep JWST images, which were then degraded to the LQ images in a similar fashion.

With the test dataset, we find that the restored images show significantly enhanced correlations with the GT images than their original LQ versions, reducing the scatters of isophotal photometry, Sersic index, and half-light radius by factors of 4.4, 3.6, and 4.7, respectively, with Pearson correlation coefficients approaching unity.
We also visually confirm that the restored images are superior in terms of resolution and noise level.
When we applied our model to real low-exposure HST images, the enhanced images also show significantly improved correlations with their multi-exposure versions, although the absence of their real GT images limits our interpretations. 

We discuss a few limitations of our model. First, the performance degrades in high noise regimes, where the background rms approaches $\mytilde10$\% of the object peak values. Second, highly correlated noise can be misinterpreted as astronomical features, leading to the manifestation of low-surface brightness features. Third, the restoration of point sources is less than optimal. Additionally, the model can create false object images from locally brightened regions caused by noise.

Although it is possible to further improve the model with larger training datasets and enhanced training strategies, we anticipate that our current Transformer-based deep learning model will prove useful for a number of scientific applications, including precision photometry, morphological analysis, and shear calibration.

\section*{Acknowledgements}
This work was supported by Institute of Information \& communications Technology Planning \& Evaluation (IITP) grant funded by the Korea government (MSIT) (No.2021-0-02068, Artificial Intelligence Innovation Hub).
M.J.J. acknowledges support for the current research from the National Research Foundation (NRF) of Korea under the programs 2022R1A2C1003130 and RS-2023-00219959.
This work is based [in part] on observations made with the NASA/ESA/CSA James Webb Space Telescope. The data were obtained from the Mikulski Archive for Space Telescopes at the Space Telescope Science Institute, which is operated by the Association of Universities for Research in Astronomy, Inc., under NASA contract NAS 5-03127 for JWST.
\vspace{5mm}
\facilities{JWST (NIRCam), HST (ACS)}

\software{numpy \citep{numpy}, scipy \citep{scipy}, matplotlib \citep{matplotlib}, astropy \citep{astropy2013, astropy2018}, photutils \citep{photutils}, SExtractor \citep{1996A&AS..117..393B}, GalSim \citep{Rowe2015galsim}, lacosmic \citep{lacosmic}, Drizzlepac\citep{hoffmann2021}}

\appendix

\section{Image Restoration Test with Blank Noise-only Images}
\begin{figure}
\includegraphics[width=1.01\linewidth]{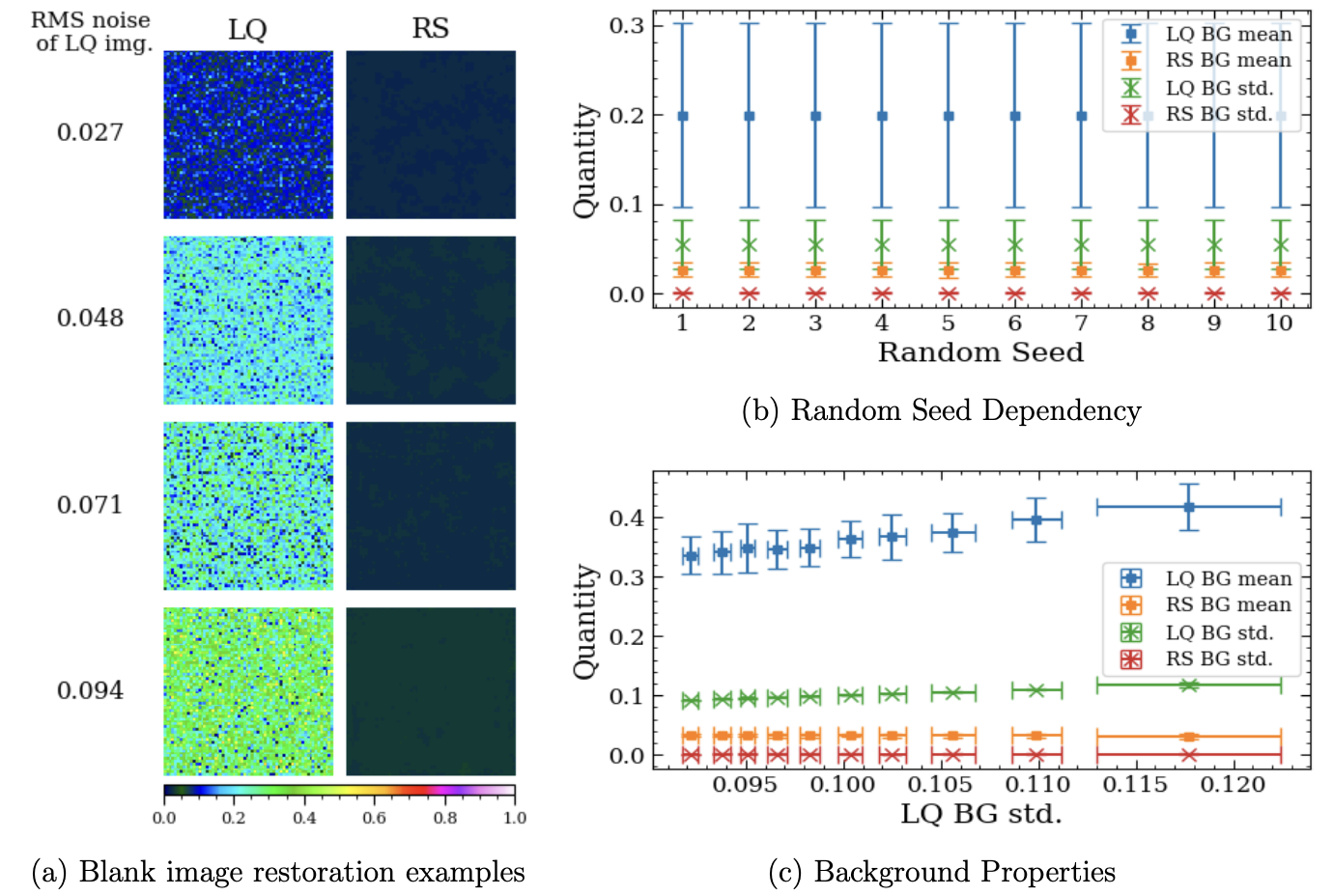}
\caption{Noise-only image restoration test.
(a) Examples of blank image restoration. The rms noise increases from top to bottom. 
No pseudo-object is produced while the noise is significantly reduced.
(b) Test with different random seeds. 
Each data point represents statistics from 1,000 images.
The mean and standard deviation of the RS images are consistent across different random seeds. 
(c) Sames as (b) but for different noise levels.  
Regardless of the input (LQ) image's noise level, the output (RS) image's background mean and noise level stay consistently low.}
\label{fig:noise_only}
\end{figure}

One of the key requirements of our deep-learning-based restoration model is that the model should not generate any false object images by overinterpreting the noise when the image contains no real astronomical source.
To test this, we created blank noise-only images by varying the random seed and the noise level. We used 10 different random seeds, and for each random seed, we generate 1000 images, where the mean and standard deviation of the noise were set to mimic those of a randomly selected galaxy image from the JWST train dataset.
The RS images created from these blank images were carefully inspected. 
No image was found to contain pseudo-sources.
Figure~\ref{fig:noise_only} presents the test results.

\bibliography{sample631}{}

\begin{thebibliography}{}
\expandafter\ifx\csname natexlab\endcsname\relax\def\natexlab#1{#1}\fi
\providecommand{\url}[1]{\href{#1}{#1}}
\providecommand{\dodoi}[1]{doi:~\href{http://doi.org/#1}{\nolinkurl{#1}}}
\providecommand{\doeprint}[1]{\href{http://ascl.net/#1}{\nolinkurl{http://ascl.net/#1}}}
\providecommand{\doarXiv}[1]{\href{https://arxiv.org/abs/#1}{\nolinkurl{https://arxiv.org/abs/#1}}}

\bibitem[{{Akhaury} {et~al.}(2022){Akhaury}, {Starck}, {Jablonka}, {Courbin}, \& {Michalewicz}}]{akhaury2022}
{Akhaury}, U., {Starck}, J.-L., {Jablonka}, P., {Courbin}, F., \& {Michalewicz}, K. 2022, Frontiers in Astronomy and Space Sciences, 9, 357, \dodoi{10.3389/fspas.2022.1001043}

\bibitem[{Alsaiari {et~al.}(2019)Alsaiari, Rustagi, Alhakamy, Thomas, \& Forbes}]{alsaiari2019}
Alsaiari, A., Rustagi, R., Alhakamy, A., Thomas, M.~M., \& Forbes, A. 2019, in 2019 IEEE 2nd International Conference on Information and Computer Technologies (ICICT), 126--132, \dodoi{10.1109/INFOCT.2019.8710893}

\bibitem[{{Astropy Collaboration} {et~al.}(2013){Astropy Collaboration}, {Robitaille}, {Tollerud}, {Greenfield}, {Droettboom}, {Bray}, {Aldcroft}, {Davis}, {Ginsburg}, {Price-Whelan}, {Kerzendorf}, {Conley}, {Crighton}, {Barbary}, {Muna}, {Ferguson}, {Grollier}, {Parikh}, {Nair}, {Unther}, {Deil}, {Woillez}, {Conseil}, {Kramer}, {Turner}, {Singer}, {Fox}, {Weaver}, {Zabalza}, {Edwards}, {Azalee Bostroem}, {Burke}, {Casey}, {Crawford}, {Dencheva}, {Ely}, {Jenness}, {Labrie}, {Lim}, {Pierfederici}, {Pontzen}, {Ptak}, {Refsdal}, {Servillat}, \& {Streicher}}]{astropy2013}
{Astropy Collaboration}, {Robitaille}, T.~P., {Tollerud}, E.~J., {et~al.} 2013, \aap, 558, A33, \dodoi{10.1051/0004-6361/201322068}

\bibitem[{{Astropy Collaboration} {et~al.}(2018){Astropy Collaboration}, {Price-Whelan}, {Sip{\H{o}}cz}, {G{\"u}nther}, {Lim}, {Crawford}, {Conseil}, {Shupe}, {Craig}, {Dencheva}, {Ginsburg}, {VanderPlas}, {Bradley}, {P{\'e}rez-Su{\'a}rez}, {de Val-Borro}, {Aldcroft}, {Cruz}, {Robitaille}, {Tollerud}, {Ardelean}, {Babej}, {Bach}, {Bachetti}, {Bakanov}, {Bamford}, {Barentsen}, {Barmby}, {Baumbach}, {Berry}, {Biscani}, {Boquien}, {Bostroem}, {Bouma}, {Brammer}, {Bray}, {Breytenbach}, {Buddelmeijer}, {Burke}, {Calderone}, {Cano Rodr{\'\i}guez}, {Cara}, {Cardoso}, {Cheedella}, {Copin}, {Corrales}, {Crichton}, {D'Avella}, {Deil}, {Depagne}, {Dietrich}, {Donath}, {Droettboom}, {Earl}, {Erben}, {Fabbro}, {Ferreira}, {Finethy}, {Fox}, {Garrison}, {Gibbons}, {Goldstein}, {Gommers}, {Greco}, {Greenfield}, {Groener}, {Grollier}, {Hagen}, {Hirst}, {Homeier}, {Horton}, {Hosseinzadeh}, {Hu}, {Hunkeler}, {Ivezi{\'c}}, {Jain}, {Jenness}, {Kanarek}, {Kendrew}, {Kern}, {Kerzendorf}, {Khvalko}, {King}, {Kirkby}, {Kulkarni},
  {Kumar}, {Lee}, {Lenz}, {Littlefair}, {Ma}, {Macleod}, {Mastropietro}, {McCully}, {Montagnac}, {Morris}, {Mueller}, {Mumford}, {Muna}, {Murphy}, {Nelson}, {Nguyen}, {Ninan}, {N{\"o}the}, {Ogaz}, {Oh}, {Parejko}, {Parley}, {Pascual}, {Patil}, {Patil}, {Plunkett}, {Prochaska}, {Rastogi}, {Reddy Janga}, {Sabater}, {Sakurikar}, {Seifert}, {Sherbert}, {Sherwood-Taylor}, {Shih}, {Sick}, {Silbiger}, {Singanamalla}, {Singer}, {Sladen}, {Sooley}, {Sornarajah}, {Streicher}, {Teuben}, {Thomas}, {Tremblay}, {Turner}, {Terr{\'o}n}, {van Kerkwijk}, {de la Vega}, {Watkins}, {Weaver}, {Whitmore}, {Woillez}, {Zabalza}, \& {Astropy Contributors}}]{astropy2018}
{Astropy Collaboration}, {Price-Whelan}, A.~M., {Sip{\H{o}}cz}, B.~M., {et~al.} 2018, \aj, 156, 123, \dodoi{10.3847/1538-3881/aabc4f}

\bibitem[{{Bertin} \& {Arnouts}(1996)}]{1996A&AS..117..393B}
{Bertin}, E., \& {Arnouts}, S. 1996, \aaps, 117, 393, \dodoi{10.1051/aas:1996164}

\bibitem[{Bradley(2023)}]{lacosmic}
Bradley, L. 2023, larrybradley/lacosmic: 1.1.0, 1.1.0,  Zenodo, \dodoi{10.5281/zenodo.10145563}

\bibitem[{{Bradley} {et~al.}(2023){Bradley}, {Sip{\H{o}}cz}, {Robitaille}, {Tollerud}, {Vin{\'\i}cius}, {Deil}, {Barbary}, {Wilson}, {Busko}, {Donath}, {G{\"u}nther}, {Cara}, {Lim}, {Me{\ss}linger}, {Conseil}, {Burnett}, {Bostroem}, {Droettboom}, {Bray}, {Andersen Bratholm}, {Jamieson}, {Ginsburg}, {Barentsen}, {Craig}, {Morris}, {Perrin}, {Rathi}, {Pascual}, {Perren}, \& {Georgiev}}]{photutils}
{Bradley}, L., {Sip{\H{o}}cz}, B., {Robitaille}, T., {et~al.} 2023, {astropy/photutils: 1.10.0}, 1.10.0,  Zenodo, \dodoi{10.5281/zenodo.1035865}

\bibitem[{{D{\'\i}az Baso} {et~al.}(2019){D{\'\i}az Baso}, {de la Cruz Rodr{\'\i}guez}, \& {Danilovic}}]{diazbaso2019}
{D{\'\i}az Baso}, C.~J., {de la Cruz Rodr{\'\i}guez}, J., \& {Danilovic}, S. 2019, \aap, 629, A99, \dodoi{10.1051/0004-6361/201936069}

\bibitem[{Elhakiem {et~al.}(2021)Elhakiem, Elsaid~Ghoniemy, \& Salama}]{elhakiem2021}
Elhakiem, A.~A., Elsaid~Ghoniemy, T., \& Salama, G.~I. 2021, in 2021 Tenth International Conference on Intelligent Computing and Information Systems (ICICIS), 51--56, \dodoi{10.1109/ICICIS52592.2021.9694140}

\bibitem[{{Ellis} {et~al.}(2013){Ellis}, {McLure}, {Dunlop}, {Robertson}, {Ono}, {Schenker}, {Koekemoer}, {Bowler}, {Ouchi}, {Rogers}, {Curtis-Lake}, {Schneider}, {Charlot}, {Stark}, {Furlanetto}, \& {Cirasuolo}}]{ellis2013}
{Ellis}, R.~S., {McLure}, R.~J., {Dunlop}, J.~S., {et~al.} 2013, \apjl, 763, L7, \dodoi{10.1088/2041-8205/763/1/L7}

\bibitem[{{Fruchter} {et~al.}(2010){Fruchter}, {Hack}, {Dencheva}, \& {Droettboom}}]{fruchter2010}
{Fruchter}, A.~S., {Hack}, W., {Dencheva}, N., \& {Droettboom}, M.~and{Greenfield}, P. 2010, in 2010 Space Telescope Science Institute Calibration Workshop, 382--387

\bibitem[{Goodfellow {et~al.}(2014)Goodfellow, Pouget-Abadie, Mirza, Xu, Warde-Farley, Ozair, Courville, \& Bengio}]{goodfellow2014generative}
Goodfellow, I., Pouget-Abadie, J., Mirza, M., {et~al.} 2014, Advances in neural information processing systems, 27

\bibitem[{{Harris} {et~al.}(2020){Harris}, {Millman}, {van der Walt}, {Gommers}, {Virtanen}, {Cournapeau}, {Wieser}, {Taylor}, {Berg}, {Smith}, {Kern}, {Picus}, {Hoyer}, {van Kerkwijk}, {Brett}, {Haldane}, {del R{\'\i}o}, {Wiebe}, {Peterson}, {G{\'e}rard-Marchant}, {Sheppard}, {Reddy}, {Weckesser}, {Abbasi}, {Gohlke}, \& {Oliphant}}]{numpy}
{Harris}, C.~R., {Millman}, K.~J., {van der Walt}, S.~J., {et~al.} 2020, \nat, 585, 357, \dodoi{10.1038/s41586-020-2649-2}

\bibitem[{{Hoffmann} {et~al.}(2021){Hoffmann}, {Mack}, {Avila}, {Martlin}, {Cohen}, \& {Bajaj}}]{hoffmann2021}
{Hoffmann}, S.~L., {Mack}, J., {Avila}, R., {et~al.} 2021, in American Astronomical Society Meeting Abstracts, Vol.~53, American Astronomical Society Meeting Abstracts, 216.02

\bibitem[{{Hunter}(2007)}]{matplotlib}
{Hunter}, J.~D. 2007, Computing in Science and Engineering, 9, 90, \dodoi{10.1109/MCSE.2007.55}

\bibitem[{{Illingworth} {et~al.}(2016){Illingworth}, {Magee}, {Bouwens}, {Oesch}, {Labbe}, {van Dokkum}, {Whitaker}, {Holden}, {Franx}, \& {Gonzalez}}]{Illingworth2016}
{Illingworth}, G., {Magee}, D., {Bouwens}, R., {et~al.} 2016, arXiv e-prints, arXiv:1606.00841, \dodoi{10.48550/arXiv.1606.00841}

\bibitem[{{Jones} \& {Wykes}(1989)}]{joneswykes1989J}
{Jones}, R., \& {Wykes}, C. 1989, {Holographic and Speckle Interferometry} (Cambridge University Press)

\bibitem[{Kalele(2023)}]{kalele2023}
Kalele, G. 2023, in 2023 1st International Conference on Innovations in High Speed Communication and Signal Processing (IHCSP), 212--215, \dodoi{10.1109/IHCSP56702.2023.10127178}

\bibitem[{{Koekemoer} {et~al.}(2013){Koekemoer}, {Ellis}, {McLure}, {Dunlop}, {Robertson}, {Ono}, {Schenker}, {Ouchi}, {Bowler}, {Rogers}, {Curtis-Lake}, {Schneider}, {Charlot}, {Stark}, {Furlanetto}, {Cirasuolo}, {Wild}, \& {Targett}}]{koekemoer2013}
{Koekemoer}, A.~M., {Ellis}, R.~S., {McLure}, R.~J., {et~al.} 2013, \apjs, 209, 3, \dodoi{10.1088/0067-0049/209/1/3}

\bibitem[{Krizhevsky {et~al.}(2012)Krizhevsky, Sutskever, \& Hinton}]{krizhevsky2012imagenet}
Krizhevsky, A., Sutskever, I., \& Hinton, G.~E. 2012, Advances in neural information processing systems, 25

\bibitem[{{Lanusse} {et~al.}(2021){Lanusse}, {Mandelbaum}, {Ravanbakhsh}, {Li}, {Freeman}, \& {P{\'o}czos}}]{Lanusse2021}
{Lanusse}, F., {Mandelbaum}, R., {Ravanbakhsh}, S., {et~al.} 2021, \mnras, 504, 5543, \dodoi{10.1093/mnras/stab1214}

\bibitem[{{Ledig} {et~al.}(2016){Ledig}, {Theis}, {Huszar}, {Caballero}, {Cunningham}, {Acosta}, {Aitken}, {Tejani}, {Totz}, {Wang}, \& {Shi}}]{ledig2016}
{Ledig}, C., {Theis}, L., {Huszar}, F., {et~al.} 2016, arXiv e-prints, arXiv:1609.04802, \dodoi{10.48550/arXiv.1609.04802}

\bibitem[{Liu {et~al.}(2021)Liu, Huang, Yu, Wang, \& Mallya}]{liu2018}
Liu, M.-Y., Huang, X., Yu, J., Wang, T.-C., \& Mallya, A. 2021, Proceedings of the IEEE, PP, 1, \dodoi{10.1109/JPROC.2021.3049196}

\bibitem[{Loshchilov \& Hutter(2018)}]{loshchilov2018fixing}
Loshchilov, I., \& Hutter, F. 2018, Fixing Weight Decay Regularization in Adam.
\newblock \url{https://openreview.net/forum?id=rk6qdGgCZ}

\bibitem[{{Lucy}(1974)}]{lucy1974}
{Lucy}, L.~B. 1974, \aj, 79, 745, \dodoi{10.1086/111605}

\bibitem[{Rajeev {et~al.}(2019)Rajeev, Samath, \& Karthikeyan}]{rajeev2019}
Rajeev, R., Samath, J., \& Karthikeyan, N. 2019, Journal of Medical Systems, 43, \dodoi{10.1007/s10916-019-1371-9}

\bibitem[{{Richardson}(1972)}]{richardson1972}
{Richardson}, W.~H. 1972, Journal of the Optical Society of America (1917-1983), 62, 55

\bibitem[{Ronneberger {et~al.}(2015)Ronneberger, Fischer, \& Brox}]{ronneberger2015u}
Ronneberger, O., Fischer, P., \& Brox, T. 2015, in Medical Image Computing and Computer-Assisted Intervention--MICCAI 2015: 18th International Conference, Munich, Germany, October 5-9, 2015, Proceedings, Part III 18, Springer, 234--241

\bibitem[{Rowe {et~al.}(2015)Rowe, Jarvis, Mandelbaum, Bernstein, Bosch, Simet, Meyers, Kacprzak, Nakajima, Zuntz, Miyatake, Dietrich, Armstrong, Melchior, \& Gill}]{Rowe2015galsim}
Rowe, B., Jarvis, M., Mandelbaum, R., {et~al.} 2015, Astronomy and Computing, 10, 121, \dodoi{10.1016/j.ascom.2015.02.002}

\bibitem[{{Schawinski} {et~al.}(2017){Schawinski}, {Zhang}, {Zhang}, {Fowler}, \& {Santhanam}}]{schawinski2017}
{Schawinski}, K., {Zhang}, C., {Zhang}, H., {Fowler}, L., \& {Santhanam}, G.~K. 2017, \mnras, 467, L110, \dodoi{10.1093/mnrasl/slx008}

\bibitem[{Shepp \& Vardi(1982)}]{sheppvardi1982}
Shepp, L.~A., \& Vardi, Y. 1982, IEEE Transactions on Medical Imaging, 1, 113, \dodoi{10.1109/TMI.1982.4307558}

\bibitem[{{Simkin}(1974)}]{simkin1974}
{Simkin}, S.~M. 1974, \aap, 31, 129

\bibitem[{{Sureau} {et~al.}(2020){Sureau}, {Lechat}, \& {Starck}}]{sureau2020}
{Sureau}, F., {Lechat}, A., \& {Starck}, J.~L. 2020, \aap, 641, A67, \dodoi{10.1051/0004-6361/201937039}

\bibitem[{{Sweere} {et~al.}(2022){Sweere}, {Valtchanov}, {Lieu}, {Vojtekova}, {Verdugo}, {Santos-Lleo}, {Pacaud}, {Briassouli}, \& {C{\'a}mpora P{\'e}rez}}]{sweere2022}
{Sweere}, S.~F., {Valtchanov}, I., {Lieu}, M., {et~al.} 2022, \mnras, 517, 4054, \dodoi{10.1093/mnras/stac2437}

\bibitem[{{Tikhonov} \& {Goncharsky}(1987)}]{tikhonov1987}
{Tikhonov}, A.~N., \& {Goncharsky}, A.~V. 1987, {Ill-posed problems in the natural sciences} (Mir Publishers)

\bibitem[{Tran {et~al.}(2021)Tran, Nguyen, \& Arai}]{tran2020}
Tran, L., Nguyen, S., \& Arai, M. 2021, GAN-Based Noise Model for Denoising Real Images (Springer, Cham), 560--572, \dodoi{10.1007/978-3-030-69538-5_34}

\bibitem[{Tripathi {et~al.}(2018)Tripathi, Lipton, \& Nguyen}]{tripath2018}
Tripathi, S., Lipton, Z., \& Nguyen, T. 2018, arXiv preprint arXiv:1803.04477

\bibitem[{{van Dokkum}(2001)}]{vandokkum2001}
{van Dokkum}, P.~G. 2001, \pasp, 113, 1420, \dodoi{10.1086/323894}

\bibitem[{Vaswani {et~al.}(2017)Vaswani, Shazeer, Parmar, Uszkoreit, Jones, Gomez, Kaiser, \& Polosukhin}]{vaswani2017}
Vaswani, A., Shazeer, N., Parmar, N., {et~al.} 2017, in Advances in neural information processing systems, 5998--6008

\bibitem[{{Virtanen} {et~al.}(2020){Virtanen}, {Gommers}, {Oliphant}, {Haberland}, {Reddy}, {Cournapeau}, {Burovski}, {Peterson}, {Weckesser}, {Bright}, {van der Walt}, {Brett}, {Wilson}, {Millman}, {Mayorov}, {Nelson}, {Jones}, {Kern}, {Larson}, {Carey}, {Polat}, {Feng}, {Moore}, {VanderPlas}, {Laxalde}, {Perktold}, {Cimrman}, {Henriksen}, {Quintero}, {Harris}, {Archibald}, {Ribeiro}, {Pedregosa}, {van Mulbregt}, \& {SciPy 1. 0 Contributors}}]{scipy}
{Virtanen}, P., {Gommers}, R., {Oliphant}, T.~E., {et~al.} 2020, Nature Methods, 17, 261, \dodoi{10.1038/s41592-019-0686-2}

\bibitem[{{Wakker} \& {Schwarz}(1988)}]{wakker1998}
{Wakker}, B.~P., \& {Schwarz}, U.~J. 1988, \aap, 200, 312

\bibitem[{Wang {et~al.}(2020)Wang, Wang, Duan, \& Li}]{wang2020}
Wang, Z., Wang, L., Duan, S., \& Li, Y. 2020, Journal of Physics: Conference Series, 1550, 032127, \dodoi{10.1088/1742-6596/1550/3/032127}

\bibitem[{{Whitaker} {et~al.}(2019){Whitaker}, {Ashas}, {Illingworth}, {Magee}, {Leja}, {Oesch}, {van Dokkum}, {Mowla}, {Bouwens}, {Franx}, {Holden}, {Labb{\'e}}, {Rafelski}, {Teplitz}, \& {Gonzalez}}]{whitaker2019}
{Whitaker}, K.~E., {Ashas}, M., {Illingworth}, G., {et~al.} 2019, \apjs, 244, 16, \dodoi{10.3847/1538-4365/ab3853}

\bibitem[{Williams \& Zipser(1989)}]{williams1989learning}
Williams, R.~J., \& Zipser, D. 1989, Neural computation, 1, 270

\bibitem[{{Yan} {et~al.}(2012){Yan}, {Fang}, \& {Zhong}}]{yan2012}
{Yan}, L., {Fang}, H., \& {Zhong}, S. 2012, Optics Letters, 37, 2778, \dodoi{10.1364/OL.37.002778}

\bibitem[{Zamir {et~al.}(2022)Zamir, Arora, Khan, Hayat, Khan, \& Yang}]{zamir2022restormer}
Zamir, S.~W., Arora, A., Khan, S., {et~al.} 2022, in Proceedings of the IEEE/CVF conference on computer vision and pattern recognition, 5728--5739

\bibitem[{{Zhang} {et~al.}(2017){Zhang}, {Zuo}, {Chen}, {Meng}, \& {Zhang}}]{zhang2017}
{Zhang}, K., {Zuo}, W., {Chen}, Y., {Meng}, D., \& {Zhang}, L. 2017, IEEE Transactions on Image Processing, 26, 3142, \dodoi{10.1109/TIP.2017.2662206}

\bibitem[{Zhang {et~al.}(2022)Zhang, Xiao, Tian, Lin, \& Zhang}]{zhang2022}
Zhang, Q., Xiao, J., Tian, C., Lin, J., \& Zhang, S. 2022, CAAI Transactions on Intelligence Technology, 8, n/a, \dodoi{10.1049/cit2.12110}

\end{thebibliography}
\bibliographystyle{aasjournal}

\end{document}